\documentclass[prd,aps,twocolumn,superscriptaddress,floatfix,preprintnumbers,nofootinbib]{revtex4-2}
\usepackage{amsmath}
\usepackage{graphicx}
\usepackage{xcolor}
\usepackage{needspace}
\usepackage{amsfonts}
\usepackage{natbib}
\usepackage{tikz}
\usepackage{siunitx}
\usetikzlibrary{math}
\usepackage{amssymb,amsmath,bm}
\usepackage[citecolor=blue,pdfa=true,linktocpage=true,urlcolor=blue,colorlinks=true]{hyperref}
\usepackage[export]{adjustbox}
\usepackage{multirow}
\usepackage{physics}
\usepackage[capitalise]{cleveref}
\usepackage{soul}
\creflabelformat{equation}{#2\textup{#1}#3}
\crefname{section}{Sec.}{Secs.}

\newcommand{\aap}{{Astron.~Astrophys.}}
\newcommand{\apjl}{{Astrophys.~J.~Lett.}}
\newcommand{\apjs}{{Astrophys.~J.~Supp.}}

\newcommand{\mnras}{{Mon.~Not.~R.~Astron.~Soc.}}
\newcommand{\jcap}{JCAP}

\newcommand{\be}{\begin{equation}}
\newcommand{\ee}{\end{equation}}

\def\ba#1\ea{\begin{align}#1\end{align}}

\newcommand{\bs}{\mathbf{s}}
\newcommand{\bk}{\mathbf{k}}

\def\bx{\mathbf{x}}

\def\bq{\mathbf{q}}

\def\br{\mathbf{r}}

\newcommand{\tj}[6]{ \begin{pmatrix}
   #1 & #2 & #3 \\
   #4 & #5 & #6 
  \end{pmatrix}}

\begin{document}

\title{Large-scale Modeling of the Observed Power Spectrum Multipoles}

\author{Robin Y. Wen}
\email{ywen@caltech.edu}

\affiliation{California Institute of Technology, 1200 East California Boulevard, Pasadena, California 91125, USA}

\author{Henry S. {Grasshorn Gebhardt}}
\affiliation{California Institute of Technology, 1200 East California Boulevard, Pasadena, California 91125, USA}
\affiliation{Jet Propulsion Laboratory, California Institute of Technology, Pasadena, California 91109, USA}

\author{Chen Heinrich}
\affiliation{California Institute of Technology, 1200 East California Boulevard, Pasadena, California 91125, USA}

\author{Olivier Dor\'e}
\affiliation{California Institute of Technology, 1200 East California Boulevard, Pasadena, California 91125, USA}
\affiliation{Jet Propulsion Laboratory, California Institute of Technology, Pasadena, California 91109, USA}

\begin{abstract}
Current and upcoming large-scale structure surveys are pushing toward increasingly wide angular coverage, where wide-angle effects—arising from the varying line of sight across the curved sky—become critical for accurate modeling of the three-dimensional galaxy power spectrum. At the same time, these survey's broader redshift reach makes the effects of redshift evolution—beyond the effective-redshift approximation—non-negligible on large radial scales. Additional observational effects such as the survey window function and integral constraints also become significant on these large scales, necessitating a careful theoretical treatment to robustly constrain local primordial non-Gaussianities and relativistic effects. In this work, we present a consistent and accurate theoretical framework for modeling the commonly used power spectrum multipoles (PSM) on large scales using the discrete spherical Fourier-Bessel (dSFB) basis. This basis ensures numerical stability and allows an exact separation between angular and radial modes. Using the dSFB basis, we study the impact of wide-angle effects and redshift evolution on the PSM, and incorporate the effects of window function convolution and integral constraints. We validate our PSM modeling using lognormal mocks under radial integral constraints with realistic survey geometries, demonstrating the readiness of our framework for application to all-sky galaxy surveys. 
\end{abstract}

\maketitle
\tableofcontents

\newpage

\section{Introduction}
The three-dimensional (3D) galaxy power spectrum in the Cartesian Fourier space is one of the central observables for Stage-IV large-scale structure surveys such as DESI \cite{24DESI_II}, Euclid \cite{24Euclid_overview}, and SPHEREx \cite{26SPHEREx} for probing fundamental physics such as the nature of inflation, dark matter, and dark energy. In an idealized scenario where the underlying comoving positions of galaxies can be measured and there is no redshift evolution, the 3D Fourier power spectrum becomes diagonal and only depends on the magnitude of the Fourier mode $k$, due to the Universe being statistically homogeneous and isotropic.   

In reality, however, galaxies are observed on our past light cone, and their measured redshifts are impacted by their peculiar velocities along the line of sight (LOS), an effect known as redshift-space distortions (RSD) \cite{87Kaiser,92Hamilton_RSD}. As a result, the Fourier power spectrum becomes anisotropic and depends on $\mu\equiv\hat{\bk}\cdot\hat{\bx}$, the angle between the galaxy's LOS and the Fourier mode direction. This angular dependence can be effectively captured by expanding the 3D Cartesian Fourier power spectrum into Legendre multipoles with respect to $\mu$, resulting in the so-called power spectrum multipoles (PSM). These PSMs can be efficiently estimated through the Yamamoto estimator \cite{06Yamamoto}, which has been widely used in 3D clustering analyses (see Refs.~\cite{17Beutler_BOSS,19Castorina_fnl_quasar,20dAmico_EFT_BOSS,20Ivanov_BOSS_EFT,21BeutlerWAWindow,24DESI_full_shape,24_DESI_Y1_PNG} for a few examples). 

In most analyses to date, the theoretical modeling of PSM under redshift-space distortions relies on the plane-parallel approximation, in which all galaxies are assumed to have the same line of sight. Under this approximation and within linear Newtonian theory, the well-known Kaiser formula \cite{87Kaiser} provides an analytical expression for PSM. While this approach is adequate for surveys with limited angular extent, it breaks down for galaxy pairs with large angular separations, as encountered in wide-field surveys. In such cases, wide-angle effects arise \cite{98Szalay_RSD_correlation_WA,15YooWA,18CastorinaWA}, capturing the deviations between the plane-parallel theoretical prediction and the observed PSM as measured by the Yamamoto estimator. These wide-angle effects can modify the PSM beyond the percent level on large scales and, if unaccounted for, can mimic a spurious local-type primordial non-Gaussianity (PNG) signal with $f_{\rm NL}\sim \mathcal{O}(1)$ \cite{23WAGR,24Benabou_WA}.  

To exactly model the wide-angle effects, Ref.~\cite{18CastorinaWA} expanded the Yamamoto estimator in terms of the spherical Fourier-Bessel (SFB) basis–product of the spherical harmonics and the spherical Bessel functions. As the eigenfunctions of the Laplacian in the
spherical coordinates, the SFB basis preserves the geometry of the curved sky and therefore naturally incorporates wide-angle effects. Ref.~\cite{18CastorinaWA} derived an analytical mapping from a generalized version of the SFB power spectrum to the PSM, which was then used in Ref.~\cite{24PSM_SFB} to exactly compute the wide-angle effects in the PSM under a full-sky window (with the geometry of a spherical shell). 

This work builds directly on Refs.~\cite{18CastorinaWA} and \cite{24PSM_SFB}, improving the formalism and incorporating observational effects that are crucial for application to survey data. For power spectrum with multipoles $L>0$, the computation of PSM in Ref.~\cite{24PSM_SFB} relies on the generalized SFB (gSFB) power spectrum, which is cumbersome and slow to calculate due to the gSFB basis being overcomplete with redundancies in both angular and radial indices (see \cref{sec:gSFB} for details). Here we switch the mapping to the discrete SFB (dSFB) basis \cite{19Samushia_SFB,21Gebhardt_SuperFab}, a proper basis decomposing a finite volume. Besides computational efficiency with fewer modes, the dSFB basis provides better numerical stability compared to the continuous one, discussed in detail in Refs.~\cite{19Samushia_SFB,24GR_SFB}. Moreover, the proper separation between angular and radial modes offered by the dSFB basis also allows easier modeling of window effects and integral constraints \cite{91Peacock,19Mattia_IC} through fast matrix multiplication, which is harder to achieve in the cSFB or gSFB basis. 

Our main result is Eq.~\eqref{eq:PSM-to-dSFB}. The exact modeling of the PSM with non-perturbative wide-angle effects and redshift evolution under window convolution has been achieved in Ref.~\cite{22CatorinaGR-P}, going through the configuration-space correlation function, while the effects of integral constraints (IC) on PSM have been studied in detail in Ref.~\cite{19Mattia_IC} under the plane-parallel limit. For the first time, we can exactly and consistently model all these effects in PSM without any approximation, and achieve fast evaluation if all mapping matrices, which do not have any cosmology dependence, are precomputed.

In order to robustly model the effects of integral constraint in the PSM, we revisit their treatment in the dSFB basis, where such effects can be handled most naturally. Building on Refs.~\cite{21Gebhardt_SuperFab,23Gebhard_SFB_eBOSS}, our formulation extends these previous results to arbitrary ICs and provides a more succinct presentation, enabling their incorporation into the PSM through Eq.~\eqref{eq:PSM-to-dSFB}.

In this work we adopt the discretized spherical Bessel functions, $g_{n\ell}(x)$ defined in Eq.~\eqref{eq:gnl}, as our choice of radial basis functions. Alternative radial bases, such as Laguerre polynomials \cite{13Leistedt_flaglet,15Leistedt_3DWL_wavelet} and Chebyshev polynomials \cite{17AngPow,24Chiarenza}, have also been employed in cosmological contexts. However, the SFB basis has a unique advantage: as the eigenfunctions of the Laplacian, it preserves the Fourier wavenumber $k$, enabling a direct connection to perturbation theory and standard Cartesian Fourier-space analyses. This accounts for the prevalence of the spherical Bessel functions over other radial functions and motivates our adoption of the dSFB basis.

Apart from the specific choice of the dSFB radial functions, we emphasize that all our analytical results—possibly with the exception of Eq.~\eqref{eq:GIC-full-sky} that explicitly relies on the velocity boundary condition of the basis—remain valid for any discretized, orthonormal radial basis. The only requirement is that the radial functions, when combined with spherical harmonics, form a complete and orthonormal basis for decomposing the 3D field.

This work is organized as follows: we review all the power spectrum statistics involved in this work in \cref{sec:PS-review} and derive the mapping from the discrete SFB power spectrum (PS) to PSM in \cref{sec:mapping}. This mapping is then used to exactly model PSM and study wide-angle effects in \cref{sec:WA} and redshift evolution in \cref{sec:RE}. We provide analytical treatment for a generic integral constraint under realistic windows in \cref{sec:IC}, and then specialize to the global and radial cases, which are directly relevant to large-scale structure surveys. We validate our theoretical calculations with measurements of PSM from lognormal simulations in \cref{sec:sim} and conclude in \cref{sec:conclusion}.

We defer some technical details and extensions of our work to the appendices. We consider the theoretical modeling of shot noise with the integral constraint in \cref{sec:shot}, where we provide both a rigorous and a heuristic derivation. We review how PSM was modeled in the plane-parallel limit in \cref{sec:standard}, which forms the baseline for comparison. In \cref{sec:detail_gsfb_to_psm}, we provide a new formulation and a detailed study on the gSFB-to-PSM mapping, building on and extending Refs.~\cite{18CastorinaWA,24PSM_SFB}. For completeness, we summarize how SFB power spectrum can be approximated by PSM, that is the reverse direction of the SFB-to-PSM mapping, in \cref{sec:psm-to-sfb}.

All numerical results presented in this work assume a best-fit Planck 2018 $\Lambda$CDM (cold dark matter) cosmology \cite{18Planck_Parameter} and a constant linear galaxy bias $b_1=1.5$ unless otherwise stated, and we assume the linear-order theory with Newtonian RSD \footnote{The Newtonian Doppler terms \cite{98Szalay_RSD_correlation_WA,10Raccanelli_WA_sim,15YooWA,24Benabou_WA} are ignored in this paper except in \cref{sec:sim} to compare with lognormal simulations, which requires the inclusion of the term.} for all calculations. We ignore the linear relativistic effects \cite{10Yoo_GR,11ChallinorLPS,11Bonvin_GR,12JeongLPS} and the observer's terms \cite{20GrimmGR,18Scaccabarozzi_GR_TP}, which we will consider in future work.

\section{Power Spectrum Statistics}\label{sec:PS-review}
\subsection{Power spectrum multipoles}
The Yamamoto estimator for measuring the galaxy power spectrum multipoles (PSM) is defined as~\cite{06Yamamoto,15Scoccimarro_FKP,15Bianchi_FFT}
\begin{equation}
\hat{P}_L^{\mathcal{O}} (\bk)\equiv\frac{2L+1}{I_{22}} \int\frac{d^2\hat{\bk}}{4\pi} F^{\mathcal{O}}_L (\bk) F^{\mathcal{O}}_0 (-\bk)\, ,
\label{eq:pl_estimator}
\end{equation}
where $I_{22}\equiv\int_{\bx} w(\bx)^2W(\bx)^2$ is the normalization factor determined by the survey window function $W(\bx)$ and the weight $w(\bx)$, and 
\begin{equation}
F^{\mathcal{O}}_L (\bk)\equiv\int_{\bx}e^{-i\bk\cdot\bx} \mathcal{L}_L(\hat{\bk}\cdot\hat{\bx})\Delta^{\mathcal{O}}(\bx)\,.
\label{eq:fl_estimator}
\end{equation}
Here $\Delta^{\mathcal{O}}(\mathbf{x})$ denotes the observed galaxy density contrast at position $\mathbf{x}$ in the configuration space. The symbol $\mathcal{O}$ emphasizes the observational effects on the field, which can contain window functions, weights, and integral constraints. We will show how the density contrast is constructed from galaxy and random catalogs in \cref{sec:general-IC}. Eq.~\eqref{eq:pl_estimator} uses the end-point LOS~\cite{19BeutlerMutipolePk}, that is choosing one of the galaxies in the galaxy pair as the LOS for the whole pair, for ease of factorization in the estimator, which enables the use of fast Fourier transforms (FFT) in the estimator~\cite{15Bianchi_FFT,15Scoccimarro_FKP,17Hand}. 

The expectation value of the Yamamoto estimator (the theoretical PSM) is then 
\begin{align}
P_{L}^{\mathcal{O}}(k)&\equiv\langle \widehat{P}^{\mathcal{O}}_L (k) \rangle = \frac{2L+1}{I_{22}} \int\frac{d^2\hat{\bk}}{4\pi}\int_{\bx_1,\bx_2} e^{-i \bk \cdot (\bx_1-\bx_2)} \nonumber\\
&\qquad \langle \Delta^{\mathcal{O}}(\bx_1)\Delta^{\mathcal{O}}(\bx_2) \rangle \mathcal{L}_L(\hat{\bx}_1\cdot \hat{\bk})\,,
\label{eq:Pl_average}
\end{align}
Our focus of this work is to accurately model Eq.~\eqref{eq:Pl_average} on the largest scales accounting for all observational effects. The modeling of Eq.~\eqref{eq:Pl_average} under the plane-parallel limit is reviewed in \cref{sec:standard}. 

In practice, the Fourier transform of Eq.~\eqref{eq:fl_estimator} is discrete up to the fundamental frequency determined by the box size, and at higher $k$ values the PS measurements at integer multiples of the fundamental frequency are typically averaged to coarser $k$ bins. To match the discrete $k$-bin choices employed by the estimator, the continuous, theoretical PSM can be integrated over the bin range $[k_i^{\rm min},k_i^{\rm max}]$
\begin{align}
   P_{L}^{\mathcal{O}}(k_i)=\frac{3}{k_{i,{\rm max}}^3-k_{i,{\rm min}}^3}\int_{k_i^{\rm min}}^{k_i^{\rm max}}dk\,k^2P^{\mathcal{O}}_L(k)\,.
   \label{eq:PSM_binned}
\end{align}

\subsection{Spherical-Fourier Bessel decomposition}
We now briefly review the SFB formalism following Ref.~\cite{21Gebhardt_SuperFab}. The spherical Fourier-Bessel decomposition expresses a field $\Delta(\bx)$ in
terms of the eigenfunctions of the Laplacian in spherical coordinates, which consist of the spherical Bessel function of the first kind $j_{\ell}(kx)$ and the spherical harmonic $Y_{\ell m}(\hat{\bx})$. We define the spherical Fourier-Bessel modes $\Delta_{\ell m}(k)$ by
\begin{align}
\Delta_{\ell m}(k) &=
\int_{\bx}j_\ell(kx)\,Y^*_{\ell m}(\hat{\bx})
\Delta(\bx)\,,
\label{eq:sfb_x-to-k}
\\
\Delta(\bx) &= \frac{2}{\pi}\int_{0}^{\infty}dk\,k^2\sum_{\ell ,m}j_\ell(kx)\,Y_{\ell m}(\hat{\bx})
\Delta_{\ell m}(k)\,,
\label{eq:sfb_k-to-x}
\end{align}
where $x$ is the comoving distance from the origin and $\hat{\bx}$ is the direction on the sky in the configuration space. The factor $2k^2/\pi$ can be split between Eqs.~\eqref{eq:sfb_x-to-k} and \eqref{eq:sfb_k-to-x} as desired\footnote{We here follow the same convention as Refs.~\cite{24PSM_SFB,24GR_SFB}, which is different from the convention adopted in Refs.~\cite{21Gebhardt_SuperFab,22Khek_SFB_fast}}. We refer to the above basis as the continuous SFB (cSFB) basis, since $k$ is a continuous quantity. 

To see the relation between the cSFB mode $\Delta_{\ell m}(k)$ and the Fourier mode $\Delta(\bk)\equiv F_{0}(\bk)$, we have (using \cref{eq:plane_wave_exp-Y,eq:SB_ortho,eq:YY_ortho})
\begin{align}
\Delta_{\ell m}(k)
&=\int_{\bx}\,j_{\ell}(kx)\,Y^*_{\ell m}(\hat{\bx})
\int\frac{d^3\bq}{(2\pi)^3}\,e^{i\bq\cdot\bx}\,\Delta(\bq)\nonumber\\
&=\frac{i^{\ell}}{4\pi}\int_{\hat{\bk}}
\,Y^*_{\ell m}(\hat{\bk})
\,\Delta(\bk)\,.
\end{align}
The above equation shows the cSFB mode as the spherical harmonic moment of the Cartesian Fourier mode in the Fourier space. This explicits the important intuition that both the Cartesian Fourier and the SFB basis separate scales according to the same wavenumber $k$, which stems from the same Laplacian eigenequation $\nabla^2 f=-k^2f$ that defines both bases, with the difference coming only from the geometry of the coordinates. The inverse of the above relation is:
\begin{align}
\label{eq:deltaklm2deltak}
\Delta(\bk)&=4\pi\sum_{\ell ,m}i^{-\ell}Y_{\ell m}(\hat{\bk})\Delta_{\ell m}(k)\,.
\end{align}

The canonical SFB power spectrum can be defined as
\begin{align}
   \langle\Delta_{\ell_1 m_1}(k_1) \Delta_{\ell_2 m_2}^*(k_2)\rangle = C_{\ell_1}(k_1,k_2)\delta_{\ell_1\ell_2}^{\rm K}\delta_{m_1m_2}^{\rm K}\,,
   \label{eq:cSFBPS-def}
\end{align} 
where the azimuthal symmetry is retained and the translational invariance is broken due to observing in the redshift space. Observational effects such as window functions and integral constraints further break the azimuthal symmetry of the field, but we can define the azimuthally averaged pseudo power spectrum
\begin{align}
C_{\ell}^{\mathcal{O}}(k_1,k_2)\equiv\frac{1}{2\ell+1}\sum_{m}\langle\Delta_{\ell m}^{\mathcal{O}}(k_1) \Delta_{\ell m}^{*\mathcal{O}}(k_2)\rangle\,.\label{eq:cSFBPS-W}
\end{align}

\subsubsection{Generalized SFB}\label{sec:gSFB}

In the above cSFB basis, the angular basis functions $Y_{\ell m}(\hat{\bx})$ and the radial basis functions $j_{\ell}(kx)$ share the same angular multipole $\ell$, whereas the generalized SFB (gSFB) decomposition allows for different $a\neq \ell$ in the radial function
\begin{align}
\Delta_{\ell m}^{a}(k) &=\int_{\bx}j_a(kx)Y^*_{\ell m}(\hat{\bx})\Delta(\bx)\,.
\label{eq:gsfb_x-to-k}
\end{align}
The generalized SFB mode was first introduced in Ref.~\cite{18CastorinaWA} to facilitate the expansion of the Cartesian PSM in the spherical coordinates (see Sec.~\ref{sec:mapping} for the mapping). It serves as a convenient intermediate quantity in our derivation toward the dSFB basis and provides continuity with earlier formulations in literature.

For each fixed radial order $a$, the generalized SFB functions are also eigenfunctions of the Laplacian, albeit with an eigenvalue of $k^2+[\ell(\ell+1)-a(a+1)]/r^2$ instead of $k^2$ for the cSFB basis. At a fixed radial order $a$, the set of functions $j_{a}(kx)Y_{\ell m}(\hat{\bx})$ for all possible $\ell,m$ and $k$ values forms a complete and non-redundant basis for a scalar field. The inverse transform of Eq.~\eqref{eq:gsfb_x-to-k} is
\begin{align}
    \Delta(\bx) &= \frac{2}{\pi}\int_{0}^{\infty}dk\,k^2j_a(kx)\,\sum_{\ell,m}Y_{\ell m}(\hat{\bx})
\Delta^a_{\ell m}(k)\,.
\label{eq:gsfb_k-to-x}
\end{align}
Using the orthogonality and completeness of spherical harmonics and spherical Bessel functions (\cref{eq:YY_ortho,eq:YY_complete,eq:SB_ortho}), one can verify that Eqs.~\eqref{eq:gsfb_x-to-k} and \eqref{eq:gsfb_k-to-x} are indeed inverse transforms of each other for a fixed $a$ value. However, if one includes all radial orders $a$ simultaneously, the union of these functions becomes over-complete. In this case, the representation of a field will become redundant, with multiple possible expansions from the combined sets of gSFB functions yielding the same $\Delta(\bx)$.

One can form the generalized SFB power spectrum
\begin{align}
   \langle\Delta_{\ell_1 m_1}^{a}(k_1) \Delta_{\ell_2 m_2}^{b*}(k_2)\rangle = C_{\ell_1}^{ab}(k_1,k_2)\delta_{\ell_1\ell_2}^{\rm K}\delta_{m_1m_2}^{\rm K}\,,
   \label{eq:gSFBPS-def}
\end{align} 
and define the azimuthally averaged pseudo power spectrum for the observed field $\Delta^{\mathcal{O}}(\bx)$ similar to Eq.~\eqref{eq:cSFBPS-W}. Compared to the cSFB power spectrum defined in Eq.~\eqref{eq:cSFBPS-def}, the generalized SFB power spectrum takes two spherical Bessel functions of orders $a$ and $b$, which can be different than the order $\ell$ of angular multipoles. When $a=b=\ell$, the generalized SFB power spectrum reduces to the usual SFB power spectrum.

\subsubsection{Discrete SFB}
The cSFB basis with a continuous $k$ value consists of eigenfunctions of the Laplacian for an infinite volume. The finite speed of light confines observations to a finite cosmic volume, and the detectability of any large-scale structure tracer is further limited to a specific redshift range, which leads to the discretization of the SFB basis. In the presence of boundary conditions at the radial direction for some finite comoving distance range $x_{\rm min}\leq x\leq x_{\rm max}$, the radial eigenfunction of the Laplacian
becomes~\cite{19Samushia_SFB,21Gebhardt_SuperFab}
\begin{align}
\label{eq:gnl}
g_{n\ell}(x) = c_{n\ell} \, j_\ell(k_{n\ell}x) + d_{n\ell} \, y_\ell(k_{n\ell}x)\,,
\end{align}
which is the linear combination of spherical Bessel functions of the first and second
kinds, solving the radial part of the Helmholtz equation shown in Eq.~\eqref{eq:helmholtz-radial}. The above coefficients are chosen to satisfy some boundary conditions and the orthonormality relation
\begin{align}
\label{eq:gnl_orthonormality}
\int_{x_{\rm min}}^{x_{\rm max}}d x\,x^2\,g_{n\ell}(x)\,g_{n'\ell}(x)
&=\delta^{\rm K}_{nn'}\,.
\end{align}
Here $n$ denotes the index for wavenumber $k_{n\ell}$ at each angular multipole $\ell$, and it represents the properly discretized radial (LOS) modes \cite{25ppn0}. In this work, we choose the velocity (Neumann) boundary condition where the derivatives of functions vanish at the boundary\footnote{See Appendix B of Ref.~\cite{25ppn0} for reasons of choosing the velocity boundary condition.}.

The dSFB decomposition of the field and its inverse transformation are
\begin{align}
\label{eq:dsfb_x-to-k}
\Delta(\bx)
&= \sum_{n,\ell,m} g_{n\ell}(x)\,Y_{\ell m}(\hat{\bx})\,\Delta_{n\ell m}\,,
\\
\label{eq:dsfb_k-to-x}
\Delta_{n \ell m}
&= \int d^3\bx\,g_{n\ell}(x)\,Y^*_{\ell m}(\hat{\bx}) \,\Delta(\bx)\,,
\end{align}
where the integration occurs over some spherical shell covering $[x_{\rm min},x_{\rm max}]$. The above transformations are the inverse of each other due to the completeness and orthogonality of the dSFB basis (\cref{eq:YY_complete,eq:dSFB-ortho,eq:dSB-complete}).

Similar to the cSFB case, one can form the dSFB power spectrum through
\begin{align}
  \langle\Delta_{n_1 \ell_1 m_1}\Delta^*_{n_2 \ell_2 m_2}\rangle =\delta_{\ell_1\ell_2}^{\rm K}\delta_{m_1m_2}^{\rm K}C_{\ell_1 n_1 n_2}\,,
\end{align}
and the pseudo power spectrum for the observed field when the azimuthal symmetry is broken
\begin{align}
C_{\ell n_1n_2}^{\mathcal{O}}\equiv\frac{1}{2\ell+1}\sum_{m}\langle\Delta_{n_1 \ell m}^{\mathcal{O}} \Delta_{n_2 \ell m}^{*\mathcal{O}}\rangle\,.\label{eq:dSFBPS-W}
\end{align}

Under the linear theory, the theoretical computation of the SFB power spectrum has been developed in Ref.~\cite{23Gebhard_SFB_eBOSS} under the linear Newtonian RSD, achieving computation of the SFB power spectrum at the order of seconds using the Iso-qr integration method\footnote{Iso-qr integration stands for 1D spherical Bessel integration along constant $q$-$r$ lines. See Appendix~F of Ref.~\cite{24GR_SFB} for a detailed description of the algorithm.}. It was later extended to include linear relativistic correction in Ref.~\cite{24GR_SFB}. The computation method applies to both the cSFB and dSFB power spectra.

\section{Mapping}\label{sec:mapping}
To express the Cartesian observables in terms of the SFB basis, Ref.~\cite{18CastorinaWA} has shown that the PSM defined in Eq.~\eqref{eq:Pl_average} can be expressed as a double sum of the gSFB power spectra (with one of the radial orders in the radial function to be the same as the angular multipole $b=\ell$)
\begin{align}
P_{L}^{\mathcal{O}}(k)&=\frac{4\pi(2L+1)}{I_{22}}\sum_{a,b}i^{-a+b}(2a+1)(2b+1)\nonumber\\
&\qquad \begin{pmatrix}
a & L & b\\
0 & 0 & 0
\end{pmatrix}^2 C_{b}^{ab,\mathcal{O}}(k,k)\,.
\label{eq:PSM-gSFB}
\end{align}
This sum is weighted by the Wigner-3$j$ symbol, indicating the geometric coupling among the PSM multipole $L$ and the gSFB radial orders $a$ and $b$.  This particular coupling coefficient is a direct result of the end-point LOS, and the geometric coupling will become significantly more complex for a generic line of sight (see \cref{sec:other-LOS}).

The Wigner-3$j$ symbol prescribes a triangle condition between $a, b$ and $L$, so for the commonly considered PSM with $L = 0$ to $4$, this sum is effectively reduced to only a few terms, making it computationally feasible. The above equation was employed in Ref.~\cite{24PSM_SFB} to exactly model the wide-angle effects in the PSM.  We will provide a different formulation of the above mapping in terms of the derivatives of spherical Bessel functions in Appendix~\ref{sec:deriv-derivation}.

In this work, we aim to build upon and improve Eq.~\eqref{eq:PSM-gSFB} by switching from the gSFB basis to the dSFB basis, which provides an exact decomposition of the field in a spherical shell. In comparison, the continuous wave number in the gSFB is redundant for observations of a finite volume. Having two spherical Bessel functions of different orders in the radial functions makes the gSFB basis overcomplete and complicates its numerical evaluation. It is also difficult to model the window convolution of the gSFB power spectrum under a generic, non-separable window. A separable window here refers to one whose angular and radial dependencies factorize, $W(\bx)=R(x)M(\hat{\bx})$, and the previous work of Ref.~\cite{24PSM_SFB} has only managed to incorporate such separable windows. As we shall see, the dSFB basis allows an exact modeling of window convolution and integral constraints under an arbitrary survey window function, enabling accurate modeling of Fourier-space clustering at the largest scales. 

\subsection{dSFB-to-gSFB mapping}\label{sec:dSFB-gSFB}

\begin{figure}[tbp]
\centerline{\includegraphics[width=0.38\textwidth]{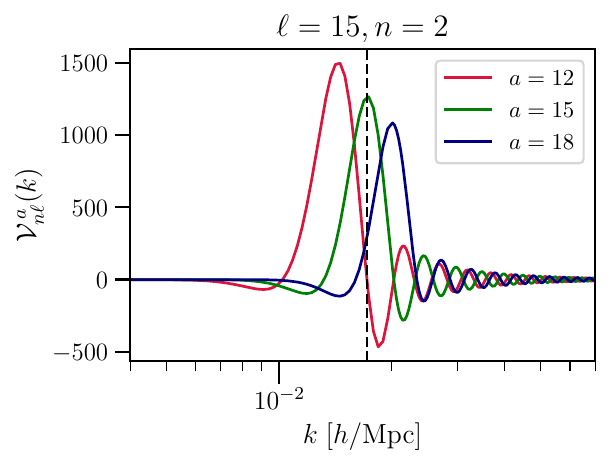}}
\caption{The dSFB-to-gSFB mapping function $\mathcal{V}_{n\ell}^{a}(k)$ with $\ell=15$ and $n=2$ at a redshift bin $z=0.2-0.5$ as defined in Eq.~\eqref{eq:d-to-g} for different radial orders $a$ in the gSFB basis. The dashed line indicates the value of $k_{n\ell}$ for the dSFB mode considered. We see that the peak of the mapping function gets shifted away from the dSFB wavenumber $k_{n\ell}$ when the gSFB radial function takes a different order than the angular multipole of the dSFB mode ($a\neq\ell$).}
\label{fig:Vnlak}
\end{figure}

To replace the gSFB power spectrum in Eq.~\eqref{eq:PSM-gSFB}, we first express the gSFB mode in terms of the dSFB mode. We consider an observed field restricted to a spherical shell spanning $[x_{\rm min}, x_{\rm max}]$, where the gSFB transform is effectively performed, and take the field $\Delta(\vec{x}) = 0$ outside this radial domain. Expanding the scalar field in Eq.~\eqref{eq:gsfb_x-to-k} in terms of the dSFB decomposition of Eq.~\eqref{eq:dsfb_x-to-k}, we obtain
\begin{align}
\Delta_{\ell m}^{a,\mathcal{O}}(k) &=
\int_{x_{\rm min}}^{x_{\rm max}}dx\,x^2
j_a(kx)\int_{\hat{\bx}}Y^*_{\ell m}(\hat{\bx})\sum_{n,\ell',m'} \nonumber\\
&\qquad g_{n\ell'}(x)\,Y_{\ell' m'}(\hat{\bx})\Delta^{\mathcal{O}}_{n\ell' m'}\nonumber\\
&=\sum_{n} \mathcal{V}^{a}_{n\ell}(k)\Delta^{\mathcal{O}}_{n\ell m}
\label{eq:d-to-g},
\end{align}
where we define the following mapping function
\begin{align}
\mathcal{V}^{a}_{n\ell}(k)\equiv\int_{x_{\rm min}}^{x_{\rm max}}dx\,x^2j_a(kx)g_{n\ell}(x)\,.
\label{eq:M_d-to-g}
\end{align}
Equation~\eqref{eq:d-to-g} gives the mapping from the dSFB mode to the gSFB mode, suggesting all gSFB modes can be constructed from the dSFB modes. The mapping function of Eq.~\eqref{eq:M_d-to-g} can be efficiently computed using the Iso-qr integration method \cite{23Gebhard_SFB_eBOSS,24GR_SFB}. We show an example of the dSFB-to-gSFB mapping function in \cref{fig:Vnlak}. 
\begin{figure}[tbp]
\centerline{\includegraphics[width=0.37\textwidth]{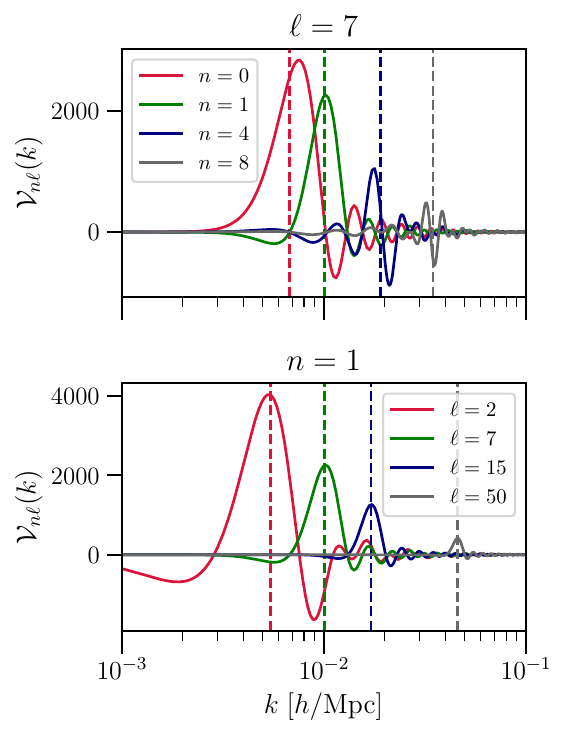}}
\caption{The dSFB-to-cSFB mapping function $\mathcal{V}_{n\ell}(k)$ as defined in Eq.~\eqref{eq:Vlnk} for different angular $\ell$ modes and radial $n$ modes at a redshift bin $z=0.2$ to $0.5$. The dashed lines indicate the value of $k_{n\ell}$ for each dSFB mode. The mapping functions peak at each of the corresponding Fourier $k_{n\ell}$ values of the dSFB mode.}
\label{fig:Vnlk}
\end{figure}

In the special case that $a=\ell$ for Eq.~\eqref{eq:d-to-g}, we obtain the mapping function from the dSFB to cSFB basis
\begin{align}
    \mathcal{V}_{n\ell}(k)\equiv\mathcal{V}^{\ell}_{n\ell}(k)\,,
    \label{eq:Vlnk}
\end{align}
which characterizes how a discretized SFB wavenumber $k_{n\ell}$ can be mapped to the continuous SFB space for a given redshift bin. We show the dSFB-to-cSFB mapping functions for various dSFB modes in \cref{fig:Vnlk}. Each mapping function exhibits a strong peak near its corresponding discrete $k_{n\ell}$ value, indicating that a given dSFB mode contributes primarily to the immediate neighborhood of the $k_{n\ell}$ value in the cSFB space. For the gSFB basis with radial order $a$ different from the angular multipoles $\ell$, the peak of the mapping functions increases (decreases) slightly  away from the corresponding $k_{n\ell}$ values for $a$ values higher (lower) than $\ell$, as shown in \cref{fig:Vnlak}.

Using Eq.~\eqref{eq:d-to-g}, all pseudo gSFB power spectra can then be evaluated as a sum of the pseudo dSFB power spectrum
\begin{align}
C_{\ell}^{ab,{\mathcal{O}}}(k_1,k_2)&=\sum_{n_1,n_2}\mathcal{V}^{a}_{n_1\ell}(k_1)\mathcal{V}^{b}_{n_2\ell}(k_2)C_{\ell n_1 n_2}^{{\mathcal{O}}}\,.
\label{eq:gSFB-to-dSFB}
\end{align}

\subsection{dSFB-to-PSM mapping}
We now obtain the mapping from the dSFB power spectrum to the Yamamoto PSM by substituting Eq.~\eqref{eq:gSFB-to-dSFB} into Eq.~\eqref{eq:PSM-gSFB}
\begin{align}
P_{L}^{\mathcal{O}}(k)&=\sum_{b,n_1,n_2}\mathcal{P}^{L}_{bn_1n_2}(k)C^{\mathcal{O}}_{bn_1n_2}\,,
\label{eq:PSM-to-dSFB}
\end{align}
where we define the following mapping function
\begin{align}
\mathcal{P}_{bn_1n_2}^{L}(k)&\equiv \frac{4\pi(2L+1)}{I_{22}}(2b+1)\mathcal{V}_{n_2b}^{b}(k)\nonumber\\
&\sum_{a}i^{-a+b}(2a+1) \begin{pmatrix}
a & L & b\\
0 & 0 & 0
\end{pmatrix}^2 \mathcal{V}_{n_1b}^{a}(k)\,.
\label{eq:dSFB-to-PSM_mapping_function}
\end{align}
Equation~\eqref{eq:PSM-to-dSFB} shows the PSM as a weighted sum of the pseudo dSFB PS, with the weights from Eq.~\eqref{eq:dSFB-to-PSM_mapping_function} completely determined by the underlying geometry transformation from the dSFB basis to the Legendre polynomials. We emphasize that $\mathcal{P}_{bn_1n_2}^{L}(k)$ only depends on the comoving distance range of the redshift bin and just needs to be computed once during cosmology inference. The mapping functions $\mathcal{P}_{bn_1n_2}^{L}(k)$ in terms of $k$ generally have similar patterns to the dSFB-to-gSFB mapping functions as shown in \cref{fig:Vnlak,fig:Vnlk}.

\begin{figure}[tbp]
\centerline{\includegraphics[width=0.45\textwidth]{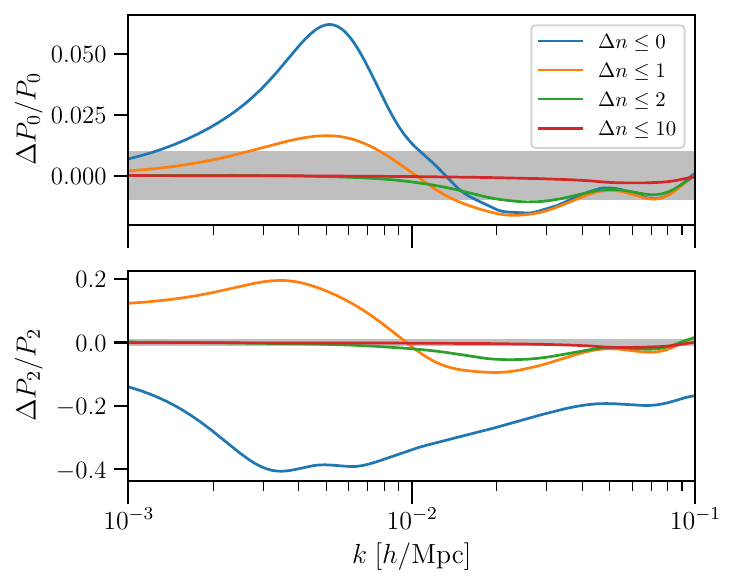}}
\caption{Diagonal and off-diagonal contribution of the dSFB power spectrum to the power spectrum monopole and quadrupole at a redshift bin $z=0.2$ to $0.5$. We show the relative deviation of the approximated PSM (including all off-diagonal contributions up to a certain $\Delta n=|n_1-n_2|$) from the true PSM. The grey band indicates the region within $1\%$ of the true PSM. The PS quadrupole has significantly more contributions from the off-diagonal SFB PS than the monopole.}
\label{fig:off_diag_contribute_02}
\end{figure}

\begin{figure*}[tbp]
\centerline{\includegraphics[width=0.86\textwidth]{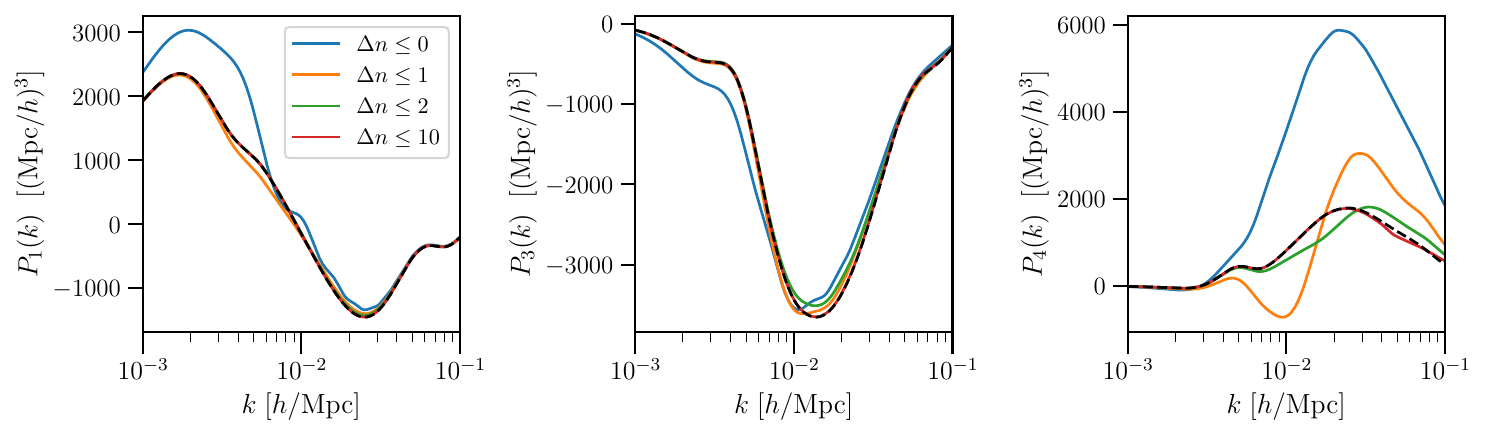}}
\caption{Power spectrum dipole, octupole, and hexadecapole obtained by summing all off-diagonal components of the dSFB power spectrum up to a certain $\Delta n$ at a redshift bin $z=0.2-0.5$. The black dashed lines indicate the correct PSM calculated with all components of the dSFB PS. We plot the absolute PSM instead of the relative error as in the case of \cref{fig:off_diag_contribute_02}, since these multipoles either cross or approach zero, making the relative error a less effective indicator of accuracy.}
\label{fig:off_diag_contribute_134}
\end{figure*}

The mapping function $\mathcal{P}_{bn_1n_2}^{L}(k)$ determines how information in the dSFB power spectrum is extracted and compressed into the PSM. For a uniform radial selection function, the diagonal components of the dSFB power spectrum are significantly larger than the off-diagonal components, implying that the PSM should be dominated by diagonal SFB terms. This expectation is confirmed for the PS monopole, dipole, and octupole, as shown in \cref{fig:off_diag_contribute_02,fig:off_diag_contribute_134}. There, we examine the relative contributions of the diagonal and off-diagonal SFB power spectrum terms to the PSM, ignoring the effects of the integral constraint and assuming a uniform spherical shell geometry in the redshift range $z=0.2$–$0.5$. For the PS monopole, diagonal contributions alone achieve percent-level accuracy (\cref{fig:off_diag_contribute_02}), indicating that it almost entirely consists of the diagonal elements. For the dipole and octupole in \cref{fig:off_diag_contribute_134}, the diagonal terms capture the overall shape and amplitude well, while for the quadrupole and hexadecapole, the off-diagonal components—especially those with $\Delta n = 1,2$—play substantial roles in the signals. In fact, the off-diagonal components constitute an increasingly larger fraction of the PSM as the multipole $L$ increases for even $L$ values as seen in \cref{fig:off_diag_contribute_02,fig:off_diag_contribute_134}.

\begin{figure}[tbp]
\centerline{\includegraphics[width=0.45\textwidth]{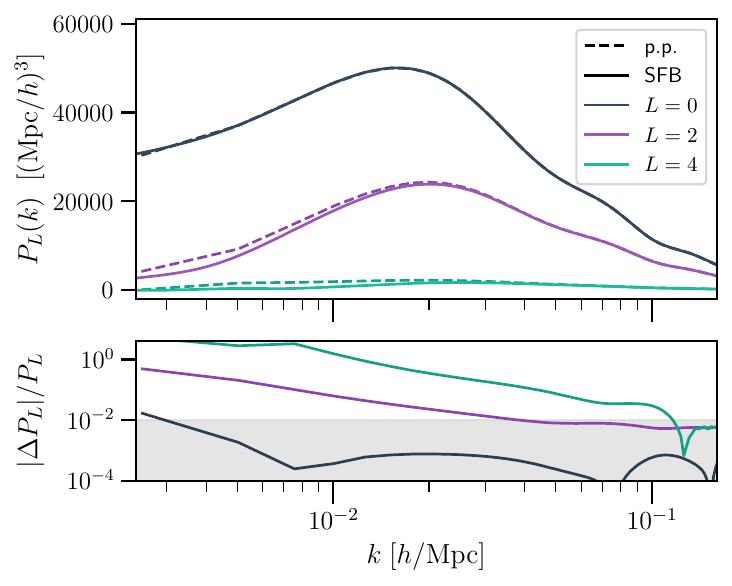}}
\caption{Comparison between the exact PSM modeled through the dSFB basis and the plane-parallel (p.p.) approximations for PS monopole, quadrupole, and hexadecapole for a uniform spherical shell of $z=0.2$ to $0.5$. The grey band in the bottom subplot indicates the region within $1\%$ of the true PSM. Here we fix the linear bias $b_1$, growth factor $D$ and growth rate $f$ to be constant across the redshift bin for both p.p. and projected SFB calculations such that we only illustrate the impact of wide-angle effects without redshift evolution. Wide-angle effects become more prominent for higher multipoles.}
\label{fig:WA_even}
\end{figure}

\begin{figure}[tbp]
\centerline{\includegraphics[width=0.375\textwidth]{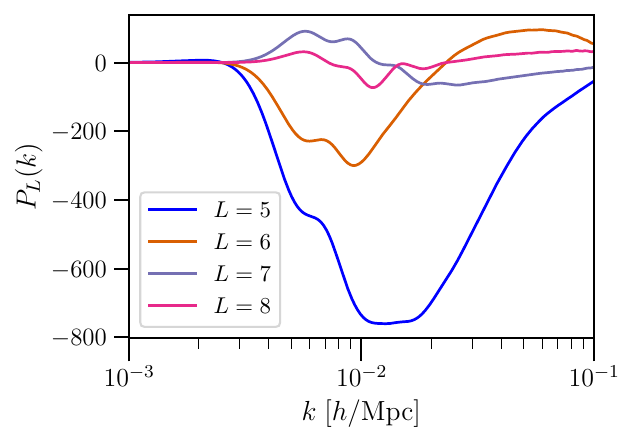}}
\caption{The PSM with higher Legendre multipoles $L\geq 5$ generated by wide-angle effects modeled through the dSFB basis for $z=0.2$ to $0.5$. The strength of the signal significantly decreases as $L$ increases. }
\label{fig:WA_highL}
\end{figure}

\subsection{Wide-angle effect}\label{sec:WA}
As a first application, we use Eq.~\eqref{eq:PSM-to-dSFB} to study wide-angle effects, which refer to the theoretical correction present in the 2-point correlation function (2CF) or PSM beyond the global plane-parallel (p.p.) approximation. Assuming all galaxies have the same LOS, the p.p. approximation yields the Kaiser formulas \cite{87Kaiser}, shown in Eqs.~\eqref{eq:kaiser} and \eqref{eq:kaiser-coeff}, for the unwindowed galaxy PSM (\cref{eq:local-PSM}) under the linear Newtonian theory. The p.p. approximation suffices for surveys with small angular coverage, but it breaks down for wide-field surveys. 

A commonly adopted approach for modeling wide-angle effects is to perturbatively expand in terms of galaxy separation for the 2CF \cite{16ReimbergWA,18CastorinaWA,19BeutlerMutipolePk,24Benabou_WA} or PSM \cite{21BeutlerWAWindow,23WAGR,23PaulWA}, with the zeroth-order term corresponding to the p.p. approximation. However, this perturbative approach does break down for large angular separation \cite{20Castorina_velocity_WA,22CatorinaGR-P,24Benabou_WA} and can bias the constraint on the local PNG parameter $f_{\rm NL}$ \cite{24Benabou_WA}, motivating the adoption of non-perturbative wide-angle modeling. The non-perturbative wide-angle correction on 2CF was first computed in Ref.~\cite{98Szalay_RSD_correlation_WA} and later reformulated either through the tripolar spherical harmonics~\cite{04Szapudi_WA,08Papai_WA} or the angular power spectrum~\cite{17Campagne_WA,18Tansella-GR}\footnote{They expand the 2CF in terms of angular power spectrum and then analytically sum over angular multipoles such that only numerical integration over $k$ is required.}. One can then non-perturbatively model the wide-angle effects in PSM by performing Hankel transforms on 2CFM (2CF Multipoles) as done in Refs.~\cite{18Tansella-GR,22CatorinaGR-P,23Foglieni}. 

In the previous Ref.~\cite{24PSM_SFB}, we have developed an alternative approach to non-perturbatively model the wide-angle effects in PSM using the gSFB power spectrum (through \cref{eq:PSM-gSFB}) and validated the method (for the Legendre multipoles $L=0$ to $4$) with lognormal mocks. Here we use the upgraded mapping of Eq.~\eqref{eq:PSM-to-dSFB} with the discrete SFB basis and focus on comparing the exact wide-angle results with the p.p. Kaiser formulas. 

In our modeling of wide-angle effects through the dSFB basis, the radial window function is inherently incorporated through the spherical geometry of the SFB basis. Consequently, for a uniform spherical shell in comoving space, no additional window convolution is required. In contrast, under the p.p formalism, window convolution due to the spherical shell must be explicitly performed on the Kaiser multipoles to enable a comparison with the dSFB-transformed results. For this, we follow the Fourier-space convolution method formulated in Ref.~\cite{21BeutlerWAWindow} (see Eq.~\eqref{eq:PL-window-eff-k} of \cref{sec:Cartesian-window} for review) and carry out the numerical calculations using the code \texttt{pypower}\footnote{\url{https://github.com/cosmodesi/pypower}} \cite{24DESI_II}.

The comparison between the exact PSM obtained through the dSFB basis and the plane-parallel approximations is shown in \cref{fig:WA_even}. We fix
the linear bias $b_1$, growth factor $D$, and growth rate $f$ to be constant across the redshift bin $z=0.2-0.5$ considered here. The wide-angle effect is at the percent level at the largest scales for power spectrum monopole, while it increases significantly at larger scales for higher Legendre multipoles, reaching tens of percent for quadrupole. The p.p. and exact results for the hexadecapole have significantly different behaviors up to $k \lessapprox 0.05\ h/{\rm Mpc}$.

Besides introducing corrections to the even Legendre multipoles of $L=0,2,4$, wide-angle effects also generate PS dipole and octupoles (shown in \cref{fig:off_diag_contribute_134}) and higher multipoles $L>4$ (shown in \cref{fig:WA_highL}), all of which do not exist under the p.p. Kaiser expressions. For a uniform spherical shell, the amplitude of the PS dipole and octupole is comparable to the PS hexadecapole. These odd multipoles, especially the cross-correlation between different tracers, are potentially measurable in the Stage-IV galaxy surveys \cite{23Bonvin_cross,24Blanco_multi}. 

The $L = 5$ multipole has roughly half the amplitude of the hexadecapole, and higher multipoles with $L > 5$ diminish even further and become practically negligible, as shown in \cref{fig:WA_highL}. This confirms the standard practice of limiting the analysis to multipoles with $L \leq 4$, since contributions from $L > 5$ can be ignored even under the exact wide-angle modeling.

\begin{figure}[tbp]
\centerline{\includegraphics[width=0.48\textwidth]{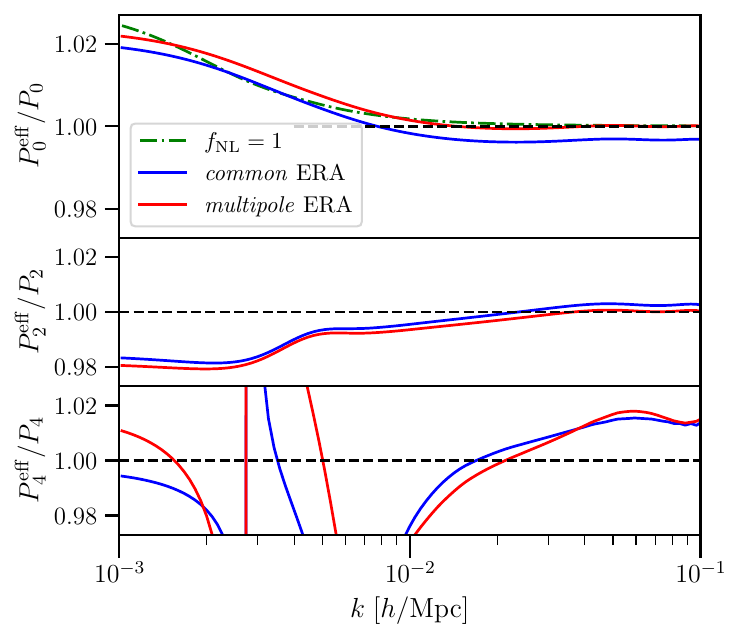}}
\caption{Comparison of the exact modeling of the PSM accounting for the evolution of growth factor $D(z)$ and growth rate $f(z)$ and the effective redshift approximations for $z=0.2$ to $0.5$. The blue line indicates the common effective redshift approximation with one effective redshift shared by all Legendre multipoles, while the red line indicates the multipole-dependent version. For the PS monopole, we also show the local PNG signal of $f_{\rm NL}=1$, which mimics the error of the effective redshift approximations at large scales.}
\label{fig:effective_redshift}
\end{figure}

\begin{figure}[tbp]
\centerline{\includegraphics[width=0.45\textwidth]{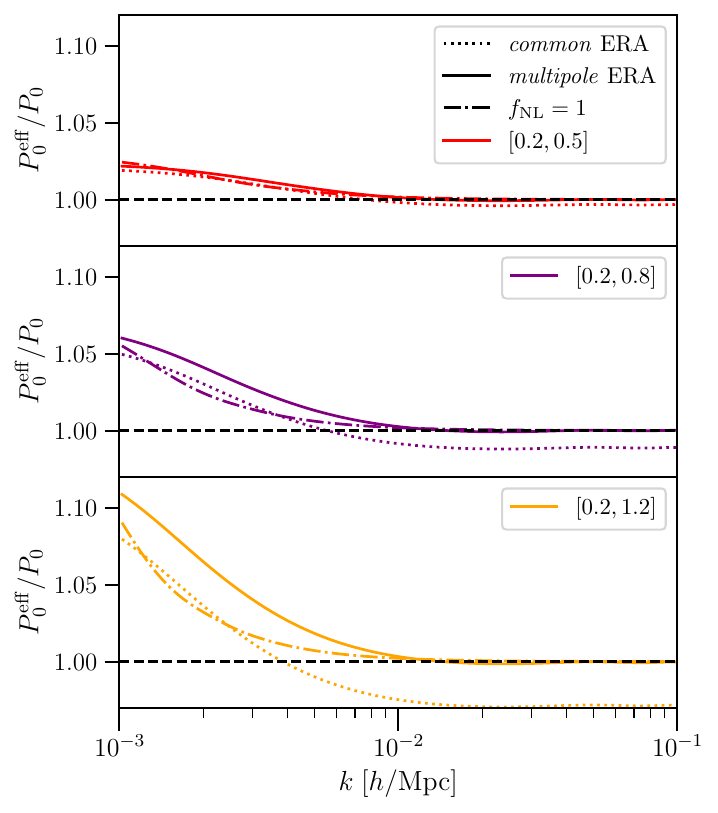}}
\caption{Comparison of the power spectrum monopole computed with the full redshift evolution to the effective redshift approximation for three redshift bins $[z_{\rm min},z_{\rm max}]$ of increasing width. Solid curves show the multipole-dependent ERA error relative to the exact calculation, dotted lines show the error for the common ERA where a single effective redshift is used for all multipoles, and dash–dot curves show the relative amplitude of a local PNG ($f_{\rm NL}=1$) signal. The ERA error grows with bin width and can exceed the amplitude of a $f_{\rm NL}=1$ signal.}
\label{fig:effective_redshift_bins}
\end{figure}

\subsection{Redshift evolution}\label{sec:RE}

We next use Eq.~\eqref{eq:PSM-to-dSFB} to exactly model the effects of redshift evolution in PSM. In the standard 3D clustering analysis, the theoretical model is evaluated at a single redshift to represent the clustering of the entire redshift bin, known as the effective redshift approximation. Such approximation is introduced to make the evaluations of 2CFM and PSM feasible for Markov chain Monte Carlo (MCMC) inference. One can use a single effective redshift for all multipoles (hereafter referred to as the common ERA), which is adopted in most 3D clustering analyses to date. Alternatively, one can define an effective redshift separately for each PS multipole to account for the anisotropic clustering due to RSD \cite{22CatorinaGR-P}
\begin{align}
    P_{L}(k,z_{{\rm eff},L})\equiv \frac{\int d^3\bx \,w(\bx)^2W(\bx)^2P_{L}(k,z(x))}{\int d^3\bx\,w(\bx)^2W(\bx)^2}\,,
    \label{eq:local_zeff-main}
\end{align}
where the multipole-dependent effective redshift $z_{{\rm eff},L}$ is defined through the weighted volume average of the PSM. We refer to this version as the multipole ERA. We review these two versions of effective redshift approximations in \cref{sec:ERA}. 

By construction, the effective redshift approximation ignores the impact of redshift evolution on the clustering signal and treats the survey volume, which spans a range of redshifts, as if it were observed at a single representative redshift. This explicit dependence of the clustering signal on the two redshifts of a galaxy pair, \(z_1\) and \(z_2\), is also referred to as the unequal-time correlation \cite{23Raccanelli_UT,23Raccanelli_AFPS}. To incorporate these effects in PSM, Refs.~\cite{23PaulWA,25Spezzati_UT} perturbatively expand the redshift dependence of cosmological and astrophysical functions in terms of galaxies’ radial separations, which captures the leading order corrections beyond the ERA.

Going through the dSFB power spectra, we can now compute the PSM with the proper evolution within the redshift bin without using the ERA or the perturbative expansion. We compare the exact results with the common and multipole ERA in \cref{fig:effective_redshift}. Here we use the dSFB basis to compute the PSM in both exact and approximate cases, corresponding to whether the growth factor $D(z)$ and growth rate $f(z)$ are allowed to vary with redshift or fixed at an effective redshift within the bin. In both cases, wide-angle effects are included such that we can isolate the effects of redshift evolution.

For power spectrum monopole and quadrupole at $z=0.2-0.5$ in \cref{fig:effective_redshift}, the multipole ERA in Eq.~\eqref{eq:local_zeff-main} produces accurate results at smaller scales of $k\geq 0.02\,h/{\rm Mpc}$, while the common version in Eq.~\eqref{eq:global_zeff} can cause subpercent-level errors by introducing a small but visible amplitude bias. Usually, this difference can be reabsorbed by changing the value of the linear galaxy bias. However, such a change is not consistent across the multipoles, and the inconsistent renormalization of galaxy bias at small scales can impact the accuracy of the $f_{\rm NL}$ or growth rate constraint. 

Therefore, both versions of effective redshift approximations remain adequate for clustering analyses excluding the ultra-large scales, while the multipole-dependent effective redshift should be preferred because it more accurately reproduces the small-scale monopole and quadrupole. In general, one can use the exact calculation results to calibrate and validate the effective redshifts needed to reproduce the clustering signals at small scales for significantly faster computations in inference.

However, at ultra-large scales, both ERAs produce percent-level inaccuracies that are scale-dependent. In the PS monopole, the error caused by ERA has a similar shape and amplitude to a local PNG signal of $f_{\rm NL}=1$, as shown in the top panel of \cref{fig:effective_redshift}. These inaccuracies grow for wider redshift bins and can even surpass a $f_{\rm NL}=1$ signal, as illustrated in \cref{fig:effective_redshift_bins}. These two figures demonstrate that relying on the ERA can bias $f_{\rm NL}$ measurements from power spectrum analyses by $\mathcal{O}(1)$. Therefore, neither version of ERA should be used in PSM calculations for surveys targeting $f_{\rm NL}\sim\mathcal{O}(1)$; a full redshift-dependent calculation should instead be carried out.

Effective redshift approximations perform significantly worse for the hexadecapole as seen in the bottom panel of \cref{fig:effective_redshift}. The multipole effective redshift is defined for the unwindowed PS hexadecapole, while we directly obtain the windowed power spectrum with a spherical-shell survey geometry through the dSFB basis. The window convolution has a strong effect on hexadecapole across scales by mixing other unwindowed multipoles, leading to the percent level inaccuracies for both ERAs at small scales. 

Our results are qualitatively in agreement with Ref.~\cite{22CatorinaGR-P}. They compute the PSM through the correlation function, while we compute it through the dSFB PS. Without employing ERA, the exact computation in 
Ref.~\cite{22CatorinaGR-P} requires non-trivial high-dimensional integration (evaluating Eq.~\eqref{eq:PL-window-standard-exact}), while our approach through the dSFB basis can be achieved in seconds for the large scales and incorporated into an MCMC analysis. 

We emphasize that both common and multipole ERA assume the galaxy selection function is isotropic in the angular direction and depends only on the redshift. This assumption is not strictly valid for realistic surveys, where the redshift distribution can vary across the sky. In such anisotropic cases, the ERA requires further modifications \cite{21Obuljen_anisotropic_ER}. In contrast, our approach treats the survey window (selection) function and the cosmological signal computation directly through the dSFB space, allowing exact window convolution without assuming the window function to be isotropic (separable) or employing ERA.

\section{Integral Constraint}\label{sec:IC}
We have so far applied the dSFB-to-PSM mapping of Eq.~\eqref{eq:PSM-to-dSFB} for modeling PSM in the idealized case of a uniform spherical shell to investigate wide-angle and redshift evolution effects. Since the dSFB-to-PSM mapping applies to the observed field, we can model observational effects in the dSFB basis and then apply the mapping to obtain the observed PSM in realistic surveys. Thanks to its discrete decomposition of the field and clear separation between angular and radial scales, the dSFB basis is well suited for modeling these observational effects. 

In this section, we focus on modeling integral constraints \cite{91Peacock,19Mattia_IC}, also known as local average effects \cite{20Wadekar_covariance,21Gebhardt_SuperFab,23Gebhard_SFB_eBOSS}, which arise because the mean galaxy number density of the field is not known but estimated from the observed field itself. Different forms of IC effects can arise depending on the specific procedure used to estimate the mean galaxy number density, with the most commonly adopted and studied cases being the global and radial ICs. Modeling these IC effects is important for measuring the local PNG through 3D galaxy clustering, since omitting them can potentially bias the $f_{\rm NL}$ constraint by $O(5)$ \cite{24_DESI_Y1_PNG}. 

In the SFB framework, earlier works such as Ref.~\cite{95Fisher_SFB} effectively accounted for IC by considering the mean density field in the SFB basis and removing the monopole components from the analysis. This procedure implicitly recognized that the global IC primarily affects the SFB monopole. More recently, Ref.~\cite{21Gebhardt_SuperFab} presented the explicit treatment of the global IC in the dSFB basis under realistic survey window functions, which was later extended to the radial IC in Ref.~\cite{23Gebhard_SFB_eBOSS}. In comparison to these two works, our formulation here generalizes to arbitrary ICs—beyond just the global and radial cases—and yields more streamlined expressions, in addition to providing more analytical intuition. Our results are consistent with these earlier results in the dSFB basis, which we discuss in \cref{sec:consistency}.

For the Cartesian Fourier PSM, the effect of radial IC is frequently modeled using a mock-based method: the mean PSM is estimated from ensembles of mocks generated with and without IC effects, and the difference is used to approximate the IC correction. Refs.~\cite{21eBOSS_quasar_fnl,24Cagliari_eBOSS_quasar_fnl} applied this strategy by comparing the means of two mock sets and then dividing the difference by the fiducial power spectrum to obtain a multiplicative correction factor. Ref.~\cite{24_DESI_Y1_PNG} further extended this approach by fitting a parametric model to the measured multiplicative correction matrix, in order to capture the anisotropic multipole mixing induced by the IC. On the analytical side, Refs.~\cite{91Peacock} and \cite{19Mattia_IC} provided treatments of global and radial ICs respectively, but only under the plane-parallel limit. In contrast, our work here provides exact analytical results for the PSM under arbitrary ICs without the plane-parallel approximation, made possible through the dSFB-to-PSM mapping. 

This section is organized as follows: the notation employed throughout this section is first introduced in \cref{sec:notation-convolution}. We provide analytical results of a generic IC in the dSFB basis in \cref{sec:general-IC}, and then consider the specific cases of global and radial ICs in \cref{sec:global-IC,sec:radial-IC}. The survey window function can be viewed as a specific form of IC, and we discuss its impact on SFB PS and PSM in \cref{sec:window}. For clarity, we consider all catalogs as continuous fields and only model the cosmological signals in the main text. The impacts of window functions, weights, and ICs on the shot noise are presented in \cref{sec:shot}.

\subsection{Notation}\label{sec:notation-convolution}
To simplify the notation of the dSFB basis functions and their associated indices, we denote
\begin{align}
\begin{split}
    e_{\mu}(\bx)&\equiv e_{n_{\mu}\ell_{\mu}m_{\mu}}(\bx)\equiv g_{n_{\mu}\ell_{\mu}}(\bx)Y^*_{\ell_{\mu}m_{\mu}}(\hat{\bx}),\\
    e^{\mu}(\bx)&\equiv e_{\mu}(\bx)^*=g_{n_{\mu}\ell_{\mu}}(\bx)Y_{\ell_{\mu}m_{\mu}}(\hat{\bx})\,,
\end{split}
\end{align}
where we use the Greek letter $\mu\equiv(n_{\mu}\ell_{\mu}m_{\mu})$ for all three dSFB indices. We use the upper and lower Greek indices to distinguish the complex conjugates of the basis functions. Such notations will also help to generalize our results in the dSFB basis to an arbitrary orthonormal basis of the field. Hereafter, we will use the two notations for SFB indices interchangeably.

A single transform corresponds to the projection of a single-variable function $f(\bx)$ onto one basis function
\be
f_{\mu}=\int d^3\bx\, e_{\mu}(\bx)  f(\bx)\label{eq:single-transform}\,,
\ee
where the integration occurs over the spherical shell that defines the dSFB basis. A double transform represents its joint projection onto a pair of basis functions
\be
f_{\rho\mu}=\int d^3\bx\, e_{\rho}(\bx)e_{\mu}(\bx) f(\bx)\label{eq:double-transform}\,,
\ee
which is symmetric between the two indices. The indices can be raised or lowered according to the basis function used in the transform.

For a bivariate function $f(\bx,\br)$ taking two vector variables, we can project to the basis function over the two arguments separately and obtain the bivariate transform
\be
f_{\rho;\mu}=\int d^3\bx \int d^3\br\, e_{\rho}(\bx)e_{\mu}(\br) f(\bx,\br)\,,
\label{eq:bivariate-transform}
\ee
where we use the semicolon to indicate the bivariate transform instead of the double transform defined in Eq.~\eqref{eq:double-transform}. Due to the orthogonality of the dSFB basis function (\cref{eq:dSFB-ortho}), the bivariate decomposition is
\begin{align}
f(\bx,\br)=\sum_{\rho,\mu}f_{\rho;\mu}e^{\rho}(\bx)e^{\mu}(\br)\,.
    \label{eq:bivariate-inverse}
\end{align}

Next, we consider the action of a general linear integral operator in the configuration space
\begin{align}
    f^{\mathcal{K}}(\bx)=\int d^3\br\,\mathcal{K}(\bx,\br)f(\br)\,,
    \label{eq:kernel}
\end{align}
where $\mathcal{K}(\bx,\br)$ is a generic kernel that may depend on both arguments. In many cosmological applications, this integral operator reduces to a standard convolution $\mathcal{K}(\bx - \br)$ or, more generally, to a non-stationary convolution $\mathcal{K}(\bx - \br, \br)$. Applying the bivariate decomposition Eq.~\eqref{eq:bivariate-inverse} on the kernel and using the orthogonality of the dSFB basis, we have
\begin{align}
    f^{\mathcal{K}}(\bx)&=\int d^3\br\,\sum_{\rho,\nu}\mathcal{K}_{\rho;}^{\phantom{ 1}\nu}e^{\rho}(\bx)e_{\nu}(\br)\sum_{\mu}e^{\mu}(\br)f_{\mu}\nonumber\\
&=\sum_{\rho}\left(\sum_{\mu}\mathcal{K}_{\rho;}^{
\phantom{1}\mu}f_{\mu}\right)e^{\rho}(\bx)\,
\end{align}
Therefore, the action of a linear integral operator in the configuration space can be expressed in the dSFB basis as matrix multiplication
\begin{align}
f^{\mathcal{K}}_{\rho}=\sum_{\mu}\mathcal{K}_{\rho;}^{\phantom{1}\mu}f_{\mu}\label{eq:convolution-dSFB}\,.
\end{align}

In the case that there is no convolution, that is $\mathcal{K}(\bx,\br)=\delta^{\rm D}(\bx-\br)$, one can verify that $\mathcal{K}_{\rho;}^{\phantom{1}\mu}=\delta^{\rm K}_{\mu\rho}$ as expected. For the case of window multiplication, that is $\mathcal{K}(\bx,\br)=W(\bx)\delta^{\rm D}(\bx-\br)$, we obtain
\begin{align}
\mathcal{K}_{\rho;}^{\phantom{1}\mu}&=\int d^3\bx\, W(\bx) e_{\rho}(\bx) \int d^3\br\, e^{\mu}(\br) \delta^{\rm D}(\bx-\br)\nonumber\\
&=W_{\rho\mu}\,
\label{eq:window-double-transform}
\end{align}
which becomes the double transform of the window function $W(\bx)$. Therefore, the double transform can be considered as a special case of the bivariate transform.

\subsection{Generic IC}\label{sec:general-IC}
With the above simplified notations, we derive the impact of a generic IC on the density fluctuation field and power spectrum starting from the basic definitions. The theoretical galaxy density contrast is defined via
\begin{align}
    \Delta(\bx)=\frac{n(\bx)-\bar{n}^{\rm true}}{\bar{n}^{\rm true}}\,,
    \label{eq:density-theory}
\end{align}
where $\bar{n}^{\rm true}$ is the expected mean density of galaxies for the whole survey footprint. However, the true ensemble-averaged number density of the galaxy field is not directly accessible, as we are limited to observation of one realization of the Universe with a finite survey volume. As a result, it must be approximated using the observed mean galaxy density, which itself contains cosmological fluctuations.

In practice, the observed density contrast is measured through \cite{94FKP,17Hand,24Benabou_WA}
\begin{align}
   \Delta^{\mathcal{O}}(\bx)&=w(\bx)\frac{n_{\rm g}(\bx)-\alpha n_{\rm s}(\bx)}{\sqrt{A}}\,.
\label{eq:density-measure}
\end{align}
where $n_{\rm g}(\bx)$ and $n_{\rm s}(\bx)$ are the number densities for the galaxy catalog and a synthetic catalog. The ratio of the number of
galaxies in the catalog to the number of objects in the synthetic catalog is denoted by $\alpha$. To improve the statistical uncertainties for the power spectrum measurements, the observed galaxy density field can be weighted, and the weights $w(\bx)$ are applied to both the data and random catalogs. For general cosmological inference, it is customary to use the Feldman–Kaiser–Peacock (FKP) weight \cite{94FKP}, which provides an approximate inverse-variance weighting, whereas alternative optimal weighting schemes are employed for measuring the local PNG \cite{19Castorina_fnl_quasar}.

The observed galaxy density is $n_{\rm g}(\bx)=W(\bx)n(\bx)$, where $W(\bx)$ is the survey window (selection) function. The synthetic catalog is generated such that it describes the expected density of observed galaxies under the survey geometry and selection in the absence of cosmological clustering
\begin{align}
    \alpha n_{\rm s}(\bx)=W(\bx)\bar{n}(\bx)\,,
    \label{eq:synthetic-catalog}
\end{align}
where $\bar{n}(\bx)$ is the estimated mean galaxy number densities, since the true mean density can not be known as discussed above. The normalization $A$ is chosen such that the expectation of the Yamamoto estimator Eq.~\eqref{eq:pl_estimator} coincides with the unwindowed PSM \cite{94FKP} 
\begin{align}
    A=\frac{1}{I_{22}}\int d^3\mathbf{x}\,w(x)^2W(x)^2\bar{n}(\bx)^2\,.
    \label{eq:density-norm}
\end{align}

The mean galaxy density can be measured through \cite{19Mattia_IC}
\begin{align}
\bar{n}(\bx)=\int d^3\br\, \epsilon(\bx,\br)W(\br)n(\br)\,,
\label{eq:measured-mean}
\end{align}
where the averaging kernel $\epsilon(\bx,\br)$ is a generic kernel characterizing the specific procedure of estimating the mean galaxy density. We note that the estimated mean density under a generic procedure can be location dependent. The normalization for the averaging kernel is required to be
\begin{align}
    1=\int d^3\br\, \epsilon(\bx,\br)W(\br)\,.
    \label{eq:IC-norm}
\end{align}
Using Eqs.~\eqref{eq:density-theory} and \eqref{eq:measured-mean}, the true mean galaxy density is related to the measured mean as
\begin{align}
    \bar{n}(\bx)=(1+\bar{\Delta}(\bx))\bar{n}^{\rm true}\,,
\end{align}
where the averaged density contrast is
\begin{align}
   \bar{\Delta}(\bx)=\int d^3\br\, \epsilon(\bx,\br)W(\br)\Delta(\br)\,.
\end{align}
Then the normalization factor becomes
\begin{align}
A&=\frac{(\bar{n}^{\rm true})^2}{I_{22}}\int d^3\mathbf{x}\,w(x)^2W(\bx)^2(1+\bar{\Delta}(\bx))^2\nonumber\\
    &=(\bar{n}^{\rm true})^2(1+O(\Delta)).
    \label{eq:estimate-A}
\end{align}

Therefore, the observed density contrast is \cite{19Mattia_IC,21Taruya_IC,23Gebhard_SFB_eBOSS}
\begin{align}
   \Delta^{\mathcal{O}}(\bx)&=w(\bx)W(\bx)\frac{n(\bx)-\bar{n}(\bx)}{\sqrt{A}}\nonumber\\
   &\simeq w(\bx)W(\bx)\left(\Delta(\bx) -\bar\Delta(\bx)\right)+O(\Delta^2)\label{eq:IC-model}\,,
\end{align}
where at the first order we can ignore the cosmological fluctuation present in the normalization factor, which is sufficient for modeling power spectrum at large scales. To simplify the notation, we absorb the window functions into the averaging kernel and define the following IC kernel
\begin{align}
    G(\bx,\br)\equiv W(\bx)\epsilon(\bx,\br)W(\br)\,.
    \label{eq:G-kernel-definition}
\end{align}
Therefore,
\begin{align}
W(\bx)\bar{\Delta}(\bx)=\int d^3\br\, G(\bx,\br)\Delta(\br)\,.
\label{eq:G-kernel}
\end{align}

Applying the results of  Eqs.~\eqref{eq:convolution-dSFB} and \eqref{eq:window-double-transform} in the dSFB basis, the observed density contrast becomes
\begin{align}
(\Delta^{\mathcal{O}})_{\rho}
&\simeq
\sum_{\mu}\left((wW)_{\rho}^{\mu}-(wG)_{\rho;}^{\phantom{x}\mu}\right)\Delta_{\mu}
\label{eq:IC-model-SFB}
\end{align}
where $(wW)_{\rho}^{\mu}$ is the double SFB transform (\cref{eq:double-transform}) of the weighted window function $(wW)(\bx)\equiv w(\bx)W(\bx)$, and $(wG)_{\rho;}^{\phantom{x}\mu}$ is the bivariate SFB transform of the weighted IC kernel $(wG)(\bx,\br)\equiv w(\bx)G(\bx,\br)$. In the dSFB space, both the impacts of window functions and IC become matrix multiplication. The full SFB power spectrum is
\begin{align}
\langle (\Delta^{\mathcal{O}})_{\rho}( \Delta^{\mathcal{O}})^{\gamma} \rangle&=\sum_{\mu,\nu}(wW-wG)_{\rho;}^{\phantom{x}\mu}\nonumber\\
&\qquad(wW-wG)^{\gamma;}_{\phantom{x}\nu}\langle \Delta_{\mu}\Delta^{\nu}\rangle\,.
\label{eq:SFB-full-PS-IC}
\end{align}

For the azimuthally averaged pseudo dSFB power spectrum defined in Eq.~\eqref{eq:dSFBPS-W}, we have the observed SFB PS as the matrix-multiplied theoretical PS
\begin{align}
 C_{\ell n_1 n_2}^{\mathcal{O}}=\sum_{L N_1 N_2} \mathcal{A}^{L N_1 N_2}_{\ell n_1n_2} C_{L N_1 N_2}\,,
 \label{eq:SFBPS-IC}
\end{align}
where the mixing matrix $\mathcal{A}$ depends on the window and IC matrices in Eq.~\eqref{eq:IC-model-SFB}
\begin{align}  \mathcal{A}&\equiv\mathcal{M}[wW-wG,wW-wG]=\mathcal{M}[wW,wW]\nonumber\\
  &-\mathcal{M}[wG,wW]-\mathcal{M}[wW,wG]+\mathcal{M}[wG,wG]\label{eq:A-mixing}\,,
\end{align}
and we use the notation $\mathcal{M}[A,B]$ to demote the azimuthally averaged contraction between two SFB matrices $A$ and $B$
\begin{align}
\mathcal{M}[A,B]_{\ell n_1 n_2}^{L N_1 N_2}\equiv\frac{1}{2\ell+1} \sum_{m,M}A_{n_1\ell m;}^{\phantom{xx}N_1LM}B^{n_2\ell m;}_{\phantom{xx}N_2LM}\,.
\label{cref:pCl-mixing}
\end{align}

\begin{figure}[tbp]
\centerline{\includegraphics[width=0.35\textwidth]{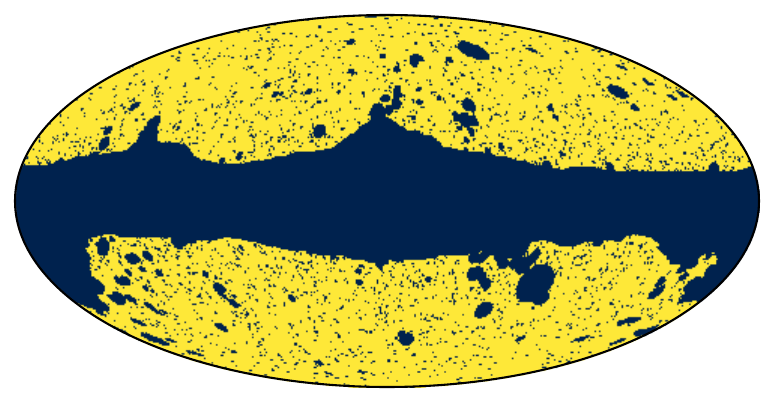}}
\caption{The angular mask for the cosmological region used in the corrected Schlegel-Finkbeiner-Davis dust map \cite{23Chiang_CSFD}, hereafter referred to as the CSFD mask. It has an effective sky fraction of $f_{\rm sky}=0.58$. This mask represents the angular footprint of galaxies for all-sky surveys such as WISE \cite{19unWISE} and SPHEREx \cite{26SPHEREx} after masking the Galactic plane and bright sources.}
\label{fig:cSFD_mask}
\end{figure}

\begin{figure*}[tbp]
\centerline{\includegraphics[width=0.8\textwidth]{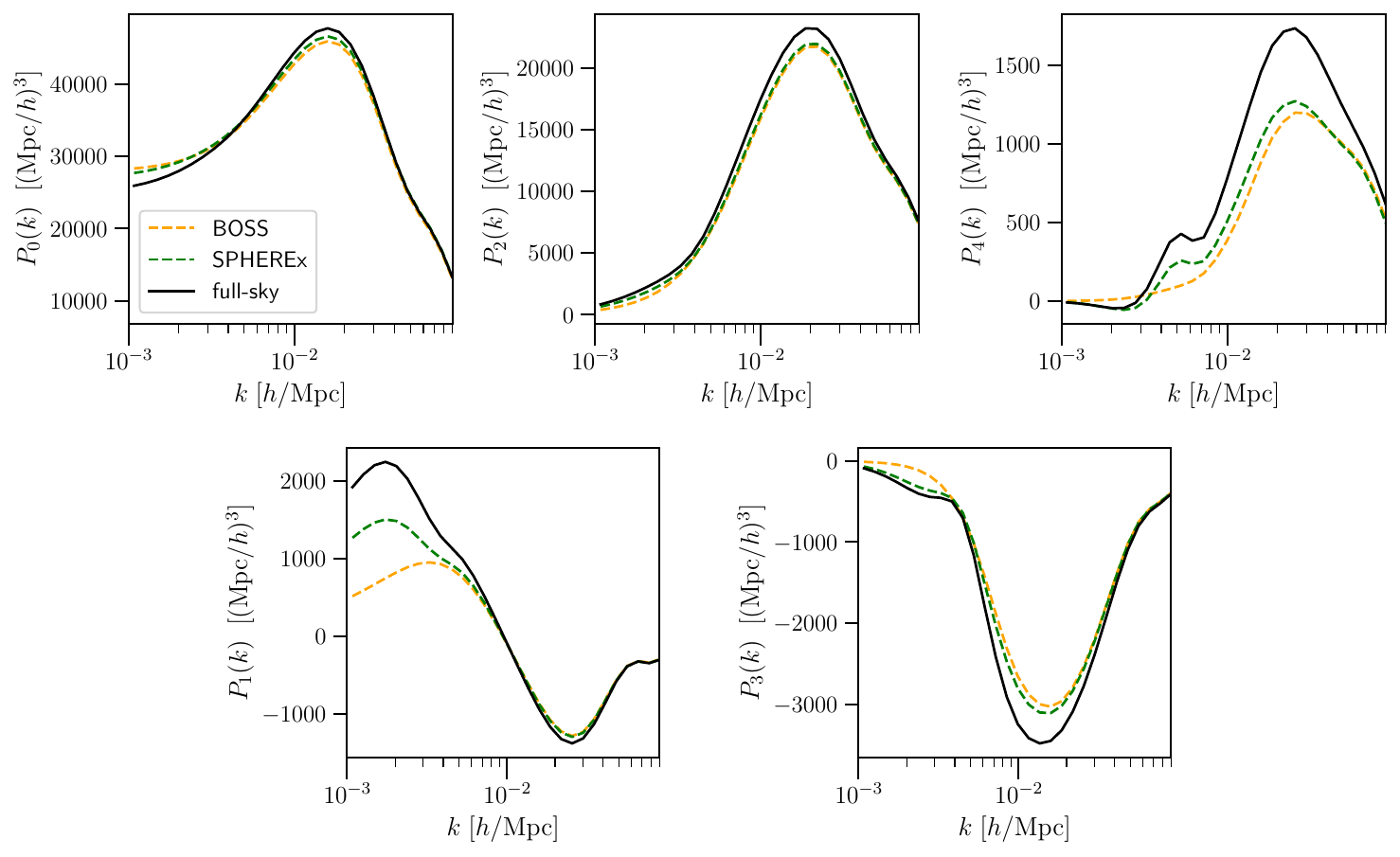}}
\caption{The power spectrum  multipoles $(L=0-4)$ in the redshift bin $z=0.2-0.5$ under the BOSS NGC, SPHEREx (CSFD), and full-sky angular masks respectively computed through the dSFB basis. The radial distribution is assumed to be uniform in all three cases.}
\label{fig:separable-window}
\end{figure*}

\subsection{Window function}\label{sec:window}
In the absence of IC effects and weights with $\Delta^{\rm W}(\bx)=W(\bx)\Delta(\bx)$, only the first term $\mathcal{M}[W,W]$ of Equation~\eqref{eq:A-mixing} remains in the mixing matrix $\mathcal{A}$. Eq.~\eqref{eq:SFBPS-IC} then gives the expression for the dSFB power spectrum under some window function, which has been studied in detail in Ref.~\cite{21Gebhardt_SuperFab} and implemented in the code \texttt{SuperFaB}. Here we use these results in the dSFB basis to model the impacts of window functions on PSM. Our approach, based on the dSFB basis, provides an alternative to the standard Cartesian-coordinate treatment of the window function (see \cref{sec:Cartesian-window} for review), and does not require the effective redshift approximation for efficient evaluation.

When the window function is the product between a radial selection function and an angular mask $W(\bx)=R(x)M(\hat{\bx})$, that is separable, the window mixing matrix in Eq.~\eqref{eq:A-mixing} greatly simplifies due to the factorization between the angular and radial parts
\ba
\mathcal{M}[W,W]_{\ell n_1n_2}^{LN_1N_2}
&=M_{\ell}^{L} R_{n_1\ell}^{N_1L}R_{n_2\ell}^{N_2L}\,,
\label{eq:separable-PS-window_matrix}
\ea
where $R_{n\ell}^{NL}$ is the double radial transform of the radial selection function
\begin{align}
    R_{n\ell}^{NL}\equiv \int d x\,x^2g_{n\ell} (x)g_{NL}(x)R(x)\,.
\label{eq:radial-coupling}
\end{align} 
$M_{\ell}^{L}$ is the angular mode coupling matrix that is commonly used to convolve the pseudo power spectrum in the cosmic microwave background and photometric galaxy surveys analyses \cite{02Hivon,19Alonso_Nmaster}
\begin{align}
M_{\ell}^{L}={\frac{(2L+1)}{4\pi}}\sum_{\ell'}
     M_{\ell'}(2\ell'+1)
    &\begin{pmatrix}
\ell & L & \ell'\\
0 & 0 & 0
\end{pmatrix}^2\,,
\label{eq:angular-mixing-window}
\end{align}
and $M_{\ell'}$ is the angular power spectrum of the angular mask $M(\hat{\bx})$. The separability of window functions significantly reduces the computational cost of the window mixing matrix $\mathcal{M}[W,W]$. 

With the separable window method, we show the first five PS multipoles under the BOSS NGC\footnote{The angular mask for the \href{https://data.sdss.org/datamodel/files/BOSSTARGET_DIR/data/geometry/boss_survey.html}{BOSS} \cite{17BOSS_DR12} North Galactic Cap (NGC) is hyperlinked.} , SPHEREx, and full-sky angular masks in \cref{fig:separable-window}. For a SPHEREx-like angular footprint, we adopt the CSFD mask \cite{23Chiang_CSFD} shown in \cref{fig:cSFD_mask}. In general, angular masks cause some suppression of signals. At small scales, the PS multipoles converge across all three cases where the impacts of angular masks are small. Overall, the PSMs under the SPHEREx-like mask track the full-sky case more closely than the BOSS case, as expected given its substantially larger angular coverage.

Our results of Eqs.~\eqref{eq:SFBPS-IC} and \eqref{eq:A-mixing} can also handle the window convolution of a generic non-separable window function. Although evaluating the double SFB transform in Eq.~\eqref{eq:double-transform} is computationally demanding, the mixing matrix only needs to be computed once, and thus does not pose a bottleneck for inference.

\subsection{Global IC}\label{sec:global-IC}
In the simplest case, we can use the average of the observed galaxy density over the whole survey volume as the estimated mean
\begin{align}
\bar{n}=\frac{1}{I_{10}}\int d^3\br\, W(\br)n(\br)\,,
\label{eq:global-IC}
\end{align}
where $I_{10}\equiv \int_{\br} W(\br)$ for the normalization. The above is known as the global IC. 

We emphasize that the global IC arises from a fundamental observational limitation: we observe only a single realization of the Universe with a finite volume. Since determining the ensemble-averaged mean galaxy number density would require knowledge of the field beyond the survey boundaries, it remains fundamentally inaccessible—making the impact of the global IC intrinsically unavoidable.

For this global case, the IC kernel defined in Eq.~\eqref{eq:G-kernel-definition} is $G(\bx,\br)=\frac{1}{I_{10}}W(\bx)W(\br)$, and the bivariate transform of the weighted IC kernel is
\begin{align}  (wG)_{\rho;}^{\phantom{x}\mu}&=\frac{1}{I_{10}}\int d^3\bx\, e_{\rho}(\bx)w(\bx)W(\bx) \int d^3\br\, e^{\mu}(\br)W(\br)\nonumber\\
    &=\frac{1}{I_{10}}(wW)_{\rho}W^{\mu}\,,
    \label{eq:GIC-SFB}
\end{align}
the outer product between the single SFB transforms of the weighted and unweighted window functions. The above mixing matrix in the dSFB basis is straightforward to construct, since $W_{\rho}$ can be efficiently evaluated. One can then obtain the mixing matrix $\mathcal{A}$ (\cref{eq:A-mixing}) for the pseudo SFB power spectrum under the global IC. 

For a full-sky uniform spherical shell without any angular masking, radial selection, or weights, the window function in the dSFB basis simplifies to
\begin{align}
    W_{\rho} =d_{\rho}\equiv\int d^3\bx\, e_{\rho}(\bx)=\sqrt{4\pi}\delta_{\ell_{\rho}0}^{\rm K}\delta^{\rm K}_{m_{\rho}0}d_{n_{\rho}0}\,,
    \label{eq:dmu}
\end{align}
where 
\begin{align}
    d_{n\ell}=\int_{x_{\rm min}}^{x_{\rm max}}dx\,x^2g_{n\ell}(x)\,.
    \label{eq:dnl}
\end{align}
Under the velocity boundary condition, the coefficient $d_{n0}$ reduces to a Kronecker delta, as shown in Eq.~\eqref{eq:d0n-identity}. Therefore,
\begin{align}
    d_{n\ell m}=\sqrt{V}\delta_{\ell0}^{\rm K}\delta^{\rm K}_{m0}\delta^{\rm K}_{n0}\,,
    \label{eq:dnlm}
\end{align}
where $V\equiv\frac{4\pi}{3}(x_{\rm max}^3-x_{\rm min}^3)$ is the full volume of the spherical shell. Under the full-sky limit, the global IC only removes the DC mode\footnote{DC mode stands for the direct current mode, by analogy with constant electrical current, referring to the spatially constant component of a field.} at the field level.

\begin{figure}[tbp]
\centerline{\includegraphics[width=0.3\textwidth]{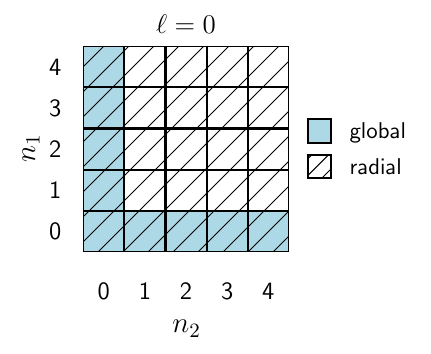}}
\caption{Components in the dSFB powers spectra removed by integral constraints for a full-sky uniform survey. We use the blue color to indicate the angular monopole modes $\ell=0$ with either $n_1=0$ or $n_2=0$ removed by the global IC, and we use diagonal shading (hatching) to highlight all the components of angular monopole $\ell=0$ removed by the radial IC.}
\label{fig:IC_mode_illustrate}
\end{figure}

Substituting Eq.~\eqref{eq:dnlm} into Eq.~\eqref{eq:GIC-SFB} and applying Eq.~\eqref{eq:A-mixing}, the dSFB power spectrum under the global IC at the full-sky limit becomes
\begin{align}
C_{\ell n_1n_2}^{\mathcal{A}}&= C_{\ell n_1n_2}-\delta^{\rm K}_{\ell 0}\delta^{\rm K}_{n_1 0}C_{00n_2}-\delta^{\rm K}_{\ell 0}\delta^{\rm K}_{n_2 0}C_{00n_1}\nonumber\\
&\quad+\delta^{\rm K}_{\ell 0}\delta^{\rm K}_{n_1 0}\delta^{\rm K}_{n_2 0}C_{000}\,,
\label{eq:GIC-full-sky}
\end{align}
where we use the symbol $\mathcal{A}$ to represent the integral constraint (local average) effect without window functions. Therefore, the global IC removes all power spectrum components containing the DC mode, that is we remove all the $\ell=0$ monopole components with either  $n_1=0$ or $n_2=0$. The last term in Eq.~\eqref{eq:GIC-full-sky} is added since the $C_{000}$ component was subtracted twice in the previous two terms. In the full-sky case, all the other angular monopole components with  $n_1\neq0$ and $n_2\neq0$ remain measurable without being removed by the global IC, as illustrated in \cref{fig:IC_mode_illustrate}.

\begin{figure*}[tbp]
\centerline{\includegraphics[width=0.8\textwidth]{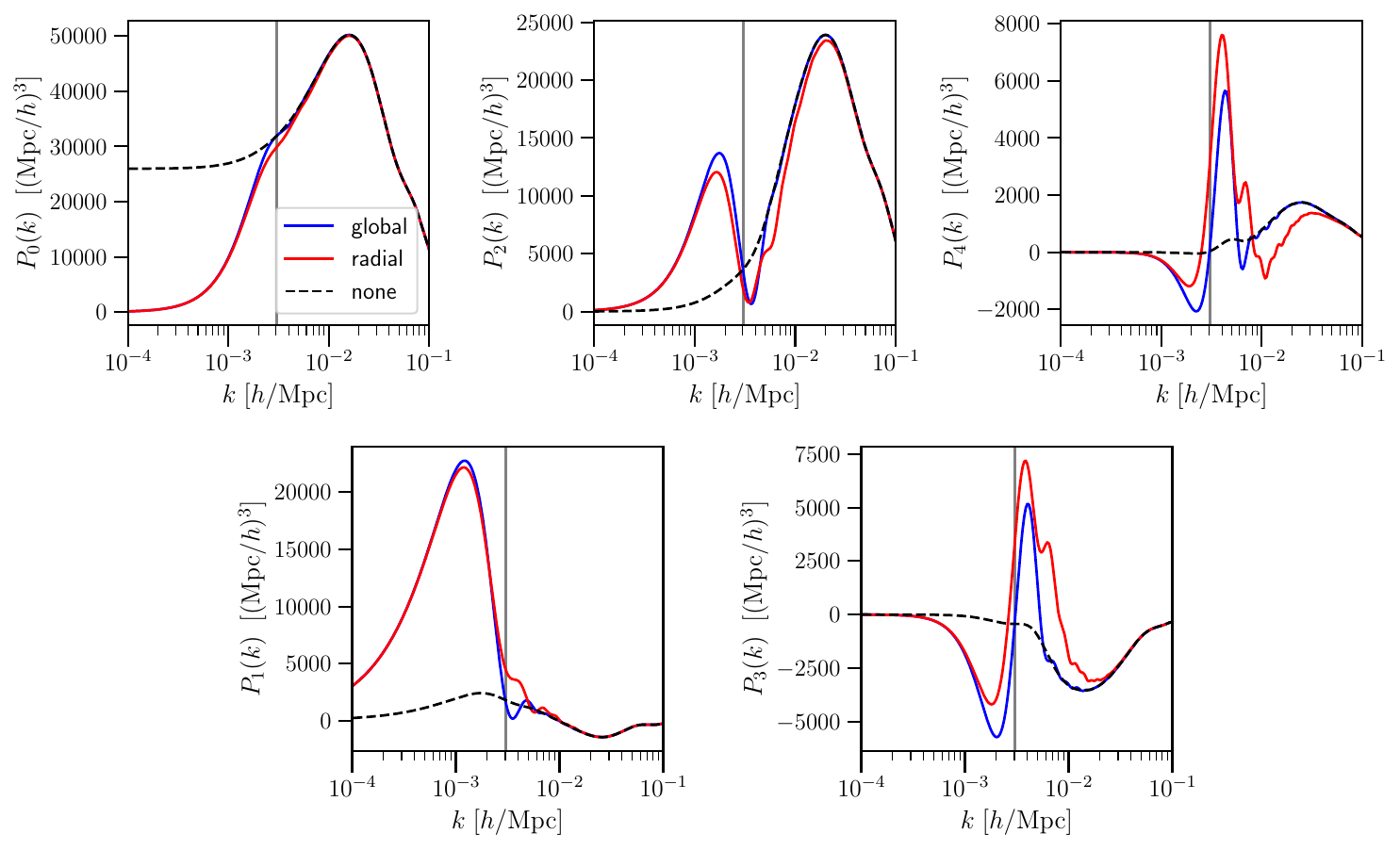}}
\caption{The power spectrum  multipoles $(L=0-4)$ under no, global, and radial integral constraints in the redshift bin $z=0.2-0.5$ with the full-sky, uniform survey geometry. The grey lines mark the survey's fundamental frequency  $k_{\rm F}\equiv 2\pi/V$. Radial integral constraints generally suppress large-scale power within the survey volume more strongly than the global case due to the removal of more angular monopole components in the dSFB PS.}
\label{fig:gic-ric-compare}
\end{figure*}

We show the PSM under the effect of global IC for the ideal full-sky geometry in \cref{fig:gic-ric-compare}. As expected, all Legendre multipoles satisfy $P_{L}(k)\to 0$ as $k\to 0$. The suppression of power near the survey's fundamental frequency is most pronounced for the Legendre monopole, and there exists significant leakage of the monopole power 
into higher multipoles compared to the no-IC case. In particular, at physical scales just a big larger than the survey volume (below the survey's fundamental frequency), the PS dipole is strongly enhanced to an amplitude comparable to the PS monopole, producing a distinctive peak.

\subsection{Radial IC}\label{sec:radial-IC}

The radial selection function of a galaxy survey often has complex dependence on the galaxy luminosity function, instrument performance, astrophysical systematics, and redshift determination algorithm. In consequence, rather than modeling the radial selection
function directly, clustering analyses usually enforce the observed galaxy redshift distribution—which contains both cosmological fluctuations and selection effects—directly onto the synthetic catalogs used for power spectrum estimation. Due to the difficulty of measuring radial selection function in practice, this approach has been adopted in all spectroscopic galaxy surveys to date (see Refs.~\cite{12Ross_BOSS_systematics,21eBOSS_quasar_fnl,24Cagliari_eBOSS_quasar_fnl,24_DESI_Y1_PNG} for examples), and its effect on the clustering statistics is known as the radial IC. 

Analytically, the observed mean number density as a function of redshift used to generate the random catalog can be expressed as
\begin{align}
\bar{n}(x)&=\frac{1}{f_{\rm eff}(x)}\int d^2\hat{\bx}\, W(\bx)n(\bx)\nonumber\\
&=\int d^3\br\,\frac{1}{x^2}\delta^{\rm D}(x-r)\widetilde{W}(\br)\Delta(\br)\,,
\label{eq:radial-IC}
\end{align}
where the second line expresses the averaging procedure at each redshift in the form of a convolution. The redshift-dependent normalization factor is $f_{\rm eff}(x)=\int d^2\hat{\bx}\, W(\bx)$ to satisfy the normalization condition in Eq.~\eqref{eq:IC-norm}, and we define the radially normalized window function $\widetilde{W}(\bx)=W(\bx)/f_{\rm eff}(x)$. Then the IC kernel defined in Eq.~\eqref{eq:G-kernel-definition} for the radial case is
\begin{align}
    G(\bx,\br)=\frac{1}{x^2}\delta^{\rm D}(x-r)W(\bx)\widetilde{W}(\br)\,.\label{eq:RIC-Gkernel}
\end{align}

Applying the completeness of $g_{n0}(x)$ functions over the index $n$ (\cref{eq:dSB-complete}), the bivariate transform of the above kernel in the dSFB basis becomes
\begin{align}
(wG)_{\rho;}^{\phantom{x}\mu}&= \sum_{n'}\int d^3\bx\,e_{\rho}(\bx)g_{n'0}(x)w(\bx) W(\bx)\nonumber\\
     &\qquad\int d^3\br\,g_{n'0}(r)  e^{\mu}(\br)\widetilde{W}(\br)\nonumber\\   &=4\pi\sum_{n'}(wW)_{n_\rho\ell_\rho m_\rho}^{n'00}\widetilde{W}^{n_{\mu}\ell_{\mu}m_{\mu}}_{n'00}\,,
     \label{eq:RIC-SFB}
\end{align}
where the last step we used $Y_{00}(\hat{\bx})=1/\sqrt{4\pi}$. For the radial IC, there is an additional radial basis function $g_{n0}(x)$ compared to the transform in the global IC case Eq.~\eqref{eq:GIC-SFB}. This leads to a double transform of the window function with one set of angular indices set to the angular monopole, which describes the coupling of the angular DC mode to higher angular multipoles due to a partial sky coverage.

For a full-sky, uniform survey with $W(\bx)=1$ in the spherical shell of $[x_{\rm min},x_{\rm max}]$, Eq.~\eqref{eq:RIC-SFB} becomes
\begin{align}
  G_{n_\rho\ell_\rho m_\rho;}^{\phantom{xx}n_\mu\ell_{\mu}m_{\mu}}&=\delta_{n_\rho n_\mu}^{\rm K}\delta_{\ell_\rho0}^{\rm K}\delta^{\rm K}_{m_\rho0}\delta_{\ell_{\mu}0}^{\rm K}\delta^{\rm K}_{m_{\mu}0}\,,
\end{align}
when we considered a uniform weight $w(\bx)=1$. One can find the three components containing IC effects in the mixing matrix $\mathcal{A}$ to be the same
\begin{align}
    \mathcal{M}[G,W]_{\ell n_1 n_2}^{LN_1N_2}&=\mathcal{M}[W,G]=\mathcal{M}[G,G]\nonumber\\
    &=\delta_{n_1N_1}^{\rm K}\delta_{n_2N_2}^{\rm K}\delta_{\ell0}^{\rm K}\delta_{L0}^{\rm K}\,.
\end{align}
Therefore, the dSFB power spectrum under the radial IC is
\begin{align}
    C_{\ell n_1 n_2}^{\mathcal{A}}=C_{\ell n_1 n_2}-\delta_{\ell 0}^{\rm K}C_{0 n_1 n_2}\,,
    \label{eq:RIC-full-sky}
\end{align}
which removes all the angular monopole components\footnote{Eq.~\eqref{eq:RIC-full-sky} recovers the result of Sec~4.4 in Ref.~\cite{22Gebhardt_crofunk}.}. Compared to the global IC case only canceling the DC mode, angular monopoles with $n_1,n_2\neq0$ are also removed due to using the measured redshift distribution as the radial selection, which we illustrate in \cref{fig:IC_mode_illustrate}. We note that Eq.~\eqref{eq:RIC-full-sky} applies to the dSFB basis under any boundary condition. 

We illustrate the PSM under the radial IC for
the ideal full-sky geometry in \cref{fig:gic-ric-compare}. We find that the effects of the radial IC are generally similar to those of the global case, but with stronger suppression of large-scale power within the survey volume, and the higher Legendre multipoles also exhibit more oscillations in the signal shape.

\begin{figure}[tbp]
\centerline{\includegraphics[width=0.4\textwidth]{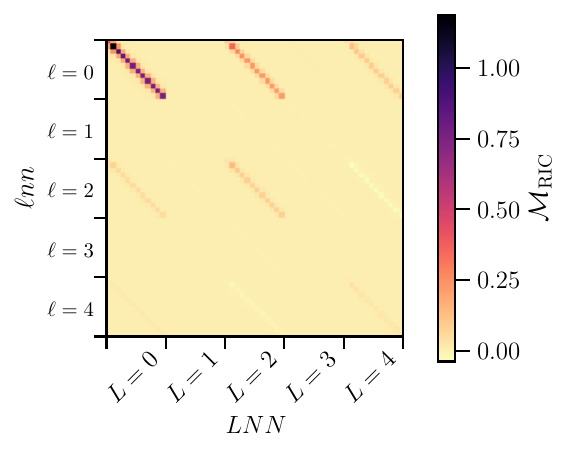}}
\caption{The radial integral constraint mixing matrix $\mathcal{M}_{\rm RIC}=\mathcal{M}[G,W]+\mathcal{M}[W,G]-\mathcal{M}[G,G]$ (as defined in Eq.~\eqref{eq:A-mixing}) for the pseudo dSFB power spectrum in the redshift bin $z=0.2$ to $0.5$ with uniform weights under the SPHEREx-like window, constructed using the CSFD angular mask and an exponentially decaying radial selection function. We show only the mixing of the diagonal components ($n_1=n_2=n$) of the dSFB PS, and the radial indices increases from left to right and from top to bottom for each angular multipole. The mixing matrix highlights the modes being canceled or suppressed by RIC. The $(LNN)$ indices denote the theoretical PS being convolved, while the $(\ell nn)$ indices denote the measured pseudo dSFB PS.}
\label{fig:RIC_SPHEREx_mixing}
\end{figure}

To introduce effects from a realistic window function, we apply the CSFD angular mask shown in \cref{fig:cSFD_mask}, representing a SPHEREx-like survey. We further impose a radial selection function of the form $\phi(r)\propto \exp(-r/r_0)$ with $r_0 = 300\,{\rm Mpc}/h$ and a normalization such that $\phi(x_{\rm min})=1$, representing the decreasing galaxy number densities observed at higher redshift due to survey depth. We will refer to this combination of angular mask and radial selection as the SPHEREx-like window hereafter.

The effects of radial IC on the dSFB power spectrum under the SPHEREx-like window is shown in \cref{fig:RIC_SPHEREx_mixing}. Similar to the full-sky case shown in \cref{fig:IC_mode_illustrate}, all angular monopole components are strongly suppressed. The window-function coupling further suppresses part of the observed angular quadrupole. Even with the SPHEREx-like window applied, the impact of the radial integral constraint remains concentrated on the largest angular scales, with negligible effects beyond $\ell=4$. The effects of RIC also concentrate on the even angular multipoles, with negligible impacts on the odd multipoles.

In principle, the radial IC can be avoided if one can accurately determine both the survey's radial selection function and the intrinsic time evolution\footnote{By ``time evolution'' we mean the change of the number density with time (redshift) for tracers selected by intrinsic properties such as absolute magnitude or host-halo mass.} of the tracer population. The radial selection function encapsulates the combined effects of instruments performance, astrophysical systematics, and analysis pipelines. It can be measured through injection simulations, where artificial sources with known redshifts and physical properties are injected into the dataset and processed through the same data pipeline in order to infer the selection and completeness of observed sources as a function of those properties. Despite the significant challenges of capturing survey complexity and scaling to modern data volumes, recent works such as Refs.~\cite{22Everett_injection,25DESILRG_injection,25DESY6_injection} have made significant progress.

Modeling the time evolution of tracer population reflects our understanding of galaxy formation. It can be approached through halo-based models or hydrodynamical simulations. Even without detailed knowledge of galaxy physics, the evolution bias $b_e$—the logarithmic derivative of the tracer's comoving number density with respect to the scale factor—can still be potentially measurable through relativistic effects in large-scale clustering \cite{15Alonso_GR_multi,19Contreras_kSZ,23Bonvin_Dipole}. 

Despite the difficulties, efforts to model or measure time evolution are essential for constraining the local PNG using two-point statistics. This is due to the equivalence between the evolution bias $b_e$ and the local PNG bias $b_\phi$ \cite{12JeongLPS,25Sullivan_bphi_be,25Dalal_bphi}. Since the PNG signal scales as $b_\phi f_{\rm NL}$, determining $b_\phi$ is critical not only for obtaining any reliable constraints on $f_{\rm NL}$ \cite{20Barreira_bphi_impact_constraint,22Barreira_bphi_BOSS} but also for enhancing them through multi-tracer analyses \cite{23Sullivan_bphi,23Barreira_multi}.

From a theoretical standpoint, the time evolution of galaxy populations is expected to vary smoothly with redshift, so ignorance of its detailed form should not impact high-frequency radial modes. In particular, small-scale radial modes (large $n$) in the dSFB basis are expected to be less sensitive to such slowly varying radial functions. This suggests that, in principle, the cosmological SFB monopoles ($\ell=0$) with large $n$ indices remain accessible to measurement for an all-sky survey, even in the presence of uncertainties in our knowledge of galaxy formation and evolution.

While the above discussion shows that enforcing radial IC may not be a necessity in principle, in practice, it is likely to be applied by default unless both the radial selection function and the time evolution of tracers can be modeled with sufficient precision. These challenges are not mere theoretical subtleties but crucial to obtaining robust and precise constraints on the local PNG. Thus, even if applying radial IC remains necessary in most realistic analyses, advancing our modeling of both radial selection and time evolution remains an important goal. In favorable situations—particularly when guided by injection simulations and relativistic clustering observables—the effects of radial IC may eventually become partially, or even fully, avoided.

\begin{figure*}[tbp]
\centerline{\includegraphics[width=0.8\textwidth]{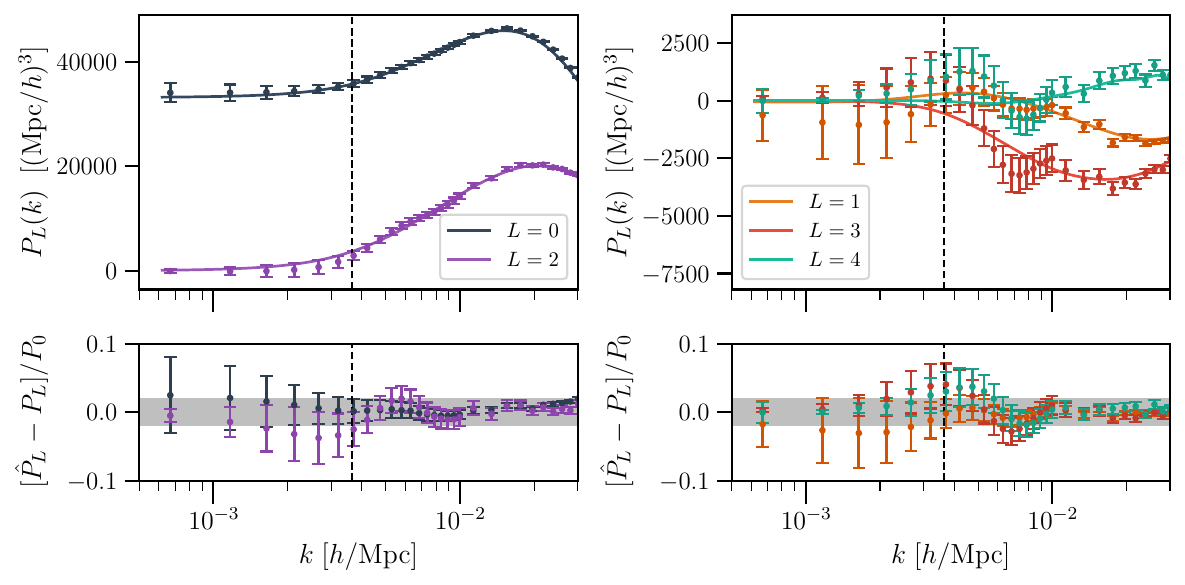}}
\caption{Theoretical and measured power spectrum multipoles $P_{L}(k)$ under the realistic SPHEREx-like window without integral constraints. The lines show the theoretical results of PSM, and the points show the mean and the $1\sigma$ error of the measured Legendre multipoles from the 1,000 lognormal mocks for $z=0.2-0.5$ based on the Yamamoto estimator. The $1\sigma$ errors shown here represent the error on the mean instead of the error for the measured spectrum from a single realization, and the errors are obtained by calculating the standard deviation of the 1,000 measured spectra. In the lower panels, we show the differences between the measurements and the theoretical results, normalized by the monopole that sets the simulation precision. The grey bands indicate $2\%$ accuracy. Our theoretical results and measurements agree well at large scales. The vertical black dashed line indicates the fundamental frequency for the effective survey volume $k_{\rm F}\equiv 2\pi/V_{\rm eff}^{1/3}$, where $V_{\rm eff}=f_{\rm sky}V$.}
\label{fig:pkl_realistic}
\end{figure*}

\begin{figure*}[tbp]
  \centering
  \begin{minipage}{0.8\textwidth}
    \centering
    \includegraphics[width=\linewidth]{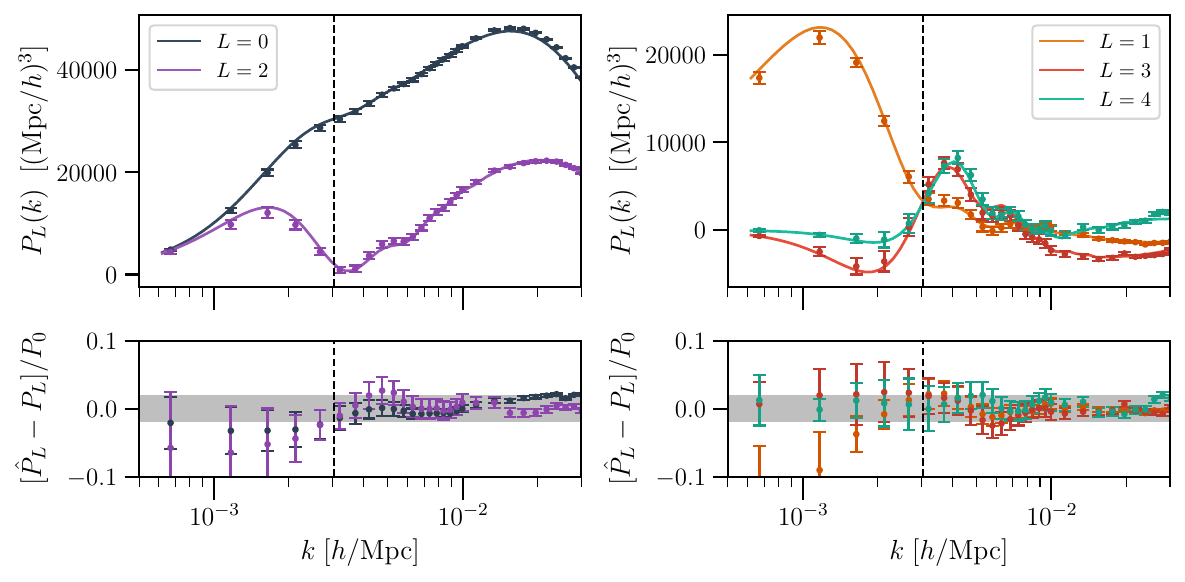}
    (a) Full-sky, radial IC
  \end{minipage}
  \par\vspace{1.5em}
  \begin{minipage}{0.8\textwidth}
    \centering
    \includegraphics[width=\linewidth]{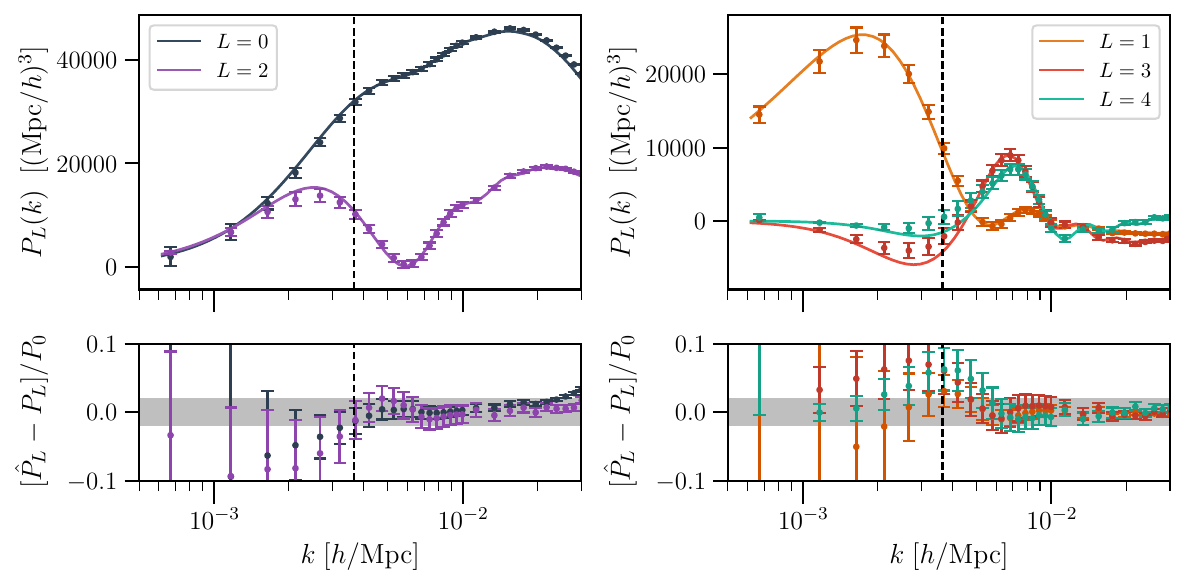}
    (b) SPHEREx-like window, radial IC
  \end{minipage}
  \caption{Validation of the theoretical modeling of power spectrum multipoles for radial integral constraints under the ideal full-sky and the realistic SPHEREx-like window cases in (a) and (b) subplots respectively. Here we adopt the same plotting convention as in \cref{fig:pkl_realistic}. We see that our theoretical predictions and means of measurements from 1,000 lognormal simulations agree well at the largest scales.}
  \label{fig:pkl_RIC_comparison}
\end{figure*}

\section{Validation with Simulations}\label{sec:sim}

To validate our large-scale PSM modeling, we compare theoretical calculations against measurements from lognormal simulations. The gSFB-to-PSM mapping has already been validated on ideal, full-sky simulations without integral constraints in Ref.~\cite{24PSM_SFB}. Here we focus on validating our modeling of the window function and radial IC effects through the dSFB-to-PSM mapping. For a realistic window, we choose a SPHEREx-like window, constructed with the CSFD angular mask and an exponentially decaying radial selection function as described in the previous section.

The lognormal mocks are generated following Ref.~\cite{24Benabou_WA}, using a reverse cloud-in-cell galaxy drawing scheme. The simulations cover the full sky area between $z=0.2$ and $0.5$, with a mean number density $\bar{n}^{\rm true}=3\times 10^{-4} h^3/{\rm Mpc}^3$. We use the matter power spectrum at an effective redshift $z_{\rm eff}=0.38$ with the Planck 2018 best-fit cosmology. We choose the linear galaxy bias $b_1=1.5$ and growth rate $f_{\rm eff}=0.71$ for the mocks, neglecting redshift evolution. 

Redshift-space distortion is implemented by generating the galaxy velocity field in linear Newtonian theory and shifting the galaxy redshifts accordingly.  Both angular mask and radial selection are applied after RSD such that the evolution bias vanishes. Besides the canonical linear Newtonian RSD term, the theoretical modeling for the lognormal mocks also needs to include the Newtonian Doppler term (also known as the $\alpha$-term) \cite{10Raccanelli_WA_sim,15YooWA,18CastorinaWA,24Benabou_WA}, which has a non-negligible impact on large scales. This term can be included in the dSFB power spectrum modeling following Refs.~\cite{24PSM_SFB,24GR_SFB}.

For measurements without IC, we use the true mean galaxy density $\bar{n}^{\rm true}$ input to generate the randoms and then apply the SPHEREx-like window function to obtain the synthetic catalogs. The random catalogs are generated at $1/\alpha=10$ times the density of the lognormal mock $\bar{n}^{\rm true}$.

To introduce the radial IC effects on the PSM measurements, we first measure the radial distribution of the lognormal mock in $N=\num{1000}$ fine redshift bins, and down-select the randoms according to the measured radial distribution of the lognormal mock, which contains both the radial selection and cosmological fluctuation. We then apply the SPHEREx-like angular mask to the down-selected random to obtain the final synthetic catalog. This procedure can be explicitly expressed as Eq.~\eqref{eq:down-selection} and reflects the redshift-dependent averaging in Eq.~\eqref{eq:radial-IC} for the radial IC. This down-selection procedure is slightly different from the shuffled method \cite{12Ross_BOSS_systematics,21Zhao_ezmock_eBOSS}, where each object in the random catalog is assigned a redshift directly drawn from the galaxy sample, but the two approaches converge in the limit of sufficiently fine binning for measuring the radial distribution of galaxies.

Power spectrum multipoles are estimated with the particular Yamamoto estimator implementation provided and validated by Ref.~\cite{24Benabou_WA}. We adopt the nearest-gridpoint (NGP) assignment scheme and corrects for both $\mu$-leakage and voxel window effects. 
To capture features at the ultra-large scales, both galaxy and random mocks are embedded in a zero-padded simulation box roughly four times the survey diameter, ensuring sensitivity to scales beyond the survey volume. This setup is particularly chosen to validate the prominent PS dipole bump caused by the radial IC at scales larger than the survey volume, as shown in \cref{fig:gic-ric-compare}.

The FFT grid fixes the native fundamental $k$-binning of the measurement, with spacing $\Delta k \simeq 5.2\times 10^{-4}\,h\,{\rm Mpc}^{-1}$ for the zero-padded simulation box. The power spectrum multipoles are measured directly on this grid. To preserve the ultra-large-scale information, we keep the native binning for $k<0.01\,h\,{\rm Mpc}^{-1}$, where the fine resolution is useful for resolving the signal shape. At higher wavenumbers, $k>0.01\,h\,{\rm Mpc}^{-1}$, we combine every four adjacent bins, giving an effective bin width $\Delta k \simeq 2.1\times 10^{-3}\,h\,{\rm Mpc}^{-1}$. This coarser binning is used for visualization to improve the readability of the plotted multipoles.

We compare our theoretical modeling of Eqs.~\eqref{eq:PSM-to-dSFB} and \eqref{eq:SFBPS-IC} against the mean of Yamamoto estimator measurements of 1,000 lognormal simulations for each configuration. PSM modes on scales comparable to or larger than the survey volume suffer from large cosmic variance and exhibit strong correlations, therefore requiring a large number of simulations to validate the modeling accuracy. For simplicity, we adopt uniform weights for all cases. To isolate the cosmological signals, we subtract the expected, theoretical shot noise from the Yamamoto measurements of lognormal mocks. The theoretical modeling of shot noise, which is impacted by ICs on large scales, is presented and validated separately in \cref{sec:shot}.

We illustrate results for the first five PS multipoles with the SPHEREx-like window without IC in \cref{fig:pkl_realistic}, which validates our modeling of window functions. We then show the effects of radial IC under both full-sky and SPHEREx-like window cases in \cref{fig:pkl_RIC_comparison}. In general, our modeling shows good agreement with the mean of simulations, even though we measure scales significantly larger than the fundamental frequency of the survey volume. The measured means agree with the theoretical prediction at the percent level within 1$\sigma$ uncertainties. 

Some residuals at the largest scales are expected from cosmic variance. Since our $k$ sampling used in the estimator is finer (larger) than the fundamental scale set by the survey volume, so adjacent PSM measurements can be strongly correlated and should not be interpreted as independent points. Therefore, for the weaker \(L=3,4\) multipoles, finite-mock fluctuations can appear as coherent oscillatory residuals around the theory curve even when the overall agreement is good.

As demonstrated in \cref{fig:pkl_RIC_comparison}, the suppression of monopole and quadrupole power with $P_{L}(k)\to 0$ as $k\to 0$ due to the radial IC effect is accurately reproduced. Moreover, radial IC significantly enhances leakage from the monopole into higher multipoles. The resulting dipole peak at super-sample scales is clearly observed in the simulations and is well captured by our theoretical predictions. Comparing the full-sky case in subplot (a) of \cref{fig:pkl_RIC_comparison} with the SPHEREx-like window case in subplot (b), we find that our theoretical modeling accurately captures the slight yet noticeable distortion of the signals induced by the SPHEREx-like window at each Legendre multipole.

\section{Conclusion and Discussion}\label{sec:conclusion}

In this work, we derived a mapping from the discrete spherical Fourier-Bessel power spectrum to the commonly used power spectrum multipoles, as presented in Eq.~\eqref{eq:PSM-to-dSFB}. This mapping connects the full-sky SFB framework with the widely adopted Fourier-based estimators, enabling precise and consistent modeling of galaxy clustering in Cartesian Fourier space at the largest scales. In particular, we demonstrated the use of this mapping for modeling the effects of wide-angle, redshift evolution, window convolution, and integral constraints in the PSM. We revisited the treatment of integral constraints in the dSFB basis, yielding a more streamlined and generalized formulation than in previous works. We further validated our PSM modeling at large-scales with measurements from lognormal simulations under realistic settings.

Beyond theoretical modeling, the dSFB-to-PSM mapping has important implications for the treatment of observational systematics in Cartesian Fourier-space analyses. Since observations are made on the past light cone, systematics often imprint characteristic patterns on angular and radial modes. However, the Yamamoto estimator for PSM mixes these scales, making it challenging to isolate and mitigate such systematics. Going through the dSFB basis, our mapping enables the removal of specific angular and radial modes directly in both the theoretical prediction and the estimator for the PSM. This allows systematics to be addressed cleanly in the dSFB basis—where their structure is often more localized \cite{25ppn0}—before being compressed into the PSM.

The radial integral constraint can be interpreted as a strategy for mitigating unknown systematics in the radial selection function. A similar approach can be applied on angular scales to suppress spurious angular fluctuations, thereby addressing unknown angular systematics. This idea, referred to as the angular integral constraint, was first introduced in Ref.~\cite{19Mattia_IC} and later applied to analysis in Ref.~\cite{24_DESI_Y1_PNG}. Our general IC results of Eqs.~\eqref{eq:SFBPS-IC} and \eqref{eq:A-mixing} can be specialized to this angular case, as well as to other forms of integral constraints that adopt different averaging procedures across different regions of the survey.

Since the dSFB basis provides a complete decomposition of the 3D field, our results can be transformed to analyze the effects of window functions and integral constraints on other two-point statistics. In particular, the angular power spectrum can be obtained by summing the dSFB power spectrum over the radial indices $n$ \cite{24GR_SFB}, while the correlation function multipoles follow from Hankel transforms of the PSM. It is also straightforward to generalize our results of the dSFB-to-PSM mapping and the modeling of integral constraints for cross-correlation among different tracers, as required in multi-tracer analyses of Stage-IV surveys.

This work highlights the utility and power of the dSFB basis for modeling of large-scale theoretical and observational effects in large-scale structure analyses. Bridging the full-sky spherical and Cartesian analyses, our mapping paves the way for robust inference in galaxy surveys targeting local primordial non-Gaussianity at $\mathcal{O}(f_{\rm NL})\lesssim 1$.

\section*{Acknowledgements} 
We thank Fynn Janssen and the rest of SPHEREx cosmology team for helpful discussion and feedback. We acknowledge support from the SPHEREx project under a contract from the NASA/GODDARD Space Flight Center to the California Institute of Technology. Part of this work was done at Jet Propulsion Laboratory, California Institute of Technology, under a contract with the National Aeronautics and Space Administration (80NM0018D0004). RYW further acknowledges support through the Canada Graduate Research Scholarship – Doctoral program (CGRS D) from the Natural Sciences and Engineering Research Council of Canada (NSERC).

\appendix

\section{Shot noise modeling} \label{sec:shot}
Real surveys are made up of discrete objects (galaxies) rather than a continuous field, which results in an additional contribution to the two-point statistics in the form of shot noise. We use $u(\bx)$ to denote a Poisson-point sample with number density $\bar{n}^{\rm true}/\alpha$ drawn from a uniform random field, that is a uniform random distribution of points. The random catalog under the survey window function is then $r(\bx)=W(\bx)u(\bx)$. We assume the galaxy catalog $n_{\rm g}(\bx)=W(\bx)n(\bx)$ is also Poisson sampled from the continuous random field. Then the auto and cross correlations of the galaxy and random catalogs are \cite{73Peebles,94FKP}
\begin{align}
    \langle n_{\rm g}(\bx)n_{\rm g}(\bx') \rangle &=\bar{n}_{\rm true}^2W(\bx)W(\bx')(1+\xi(\bx,\bx'))\nonumber\\
    &\quad+\bar{n}_{\rm true}W(\bx)\delta^{\rm D}(\bx-\bx')\,
    \label{eq:Poisson-shot-noise}\\
     \langle r(\bx)r(\bx') \rangle &=\alpha^{-2}\bar{n}_{\rm true}^2W(\bx)W(\bx')\nonumber\\
    &\quad+\alpha^{-1}\bar{n}_{\rm true}W(\bx)\delta^{\rm D}(\bx-\bx')
    \label{eq:random-shot-noise}\\
    \langle n_{\rm g}(\bx)r(\bx') \rangle &=\alpha^{-1}\bar{n}_{\rm true}^2W(\bx)W(\bx')\,,
    \label{eq:random-galaxy cross}
\end{align}
where $\xi(\bx,\bx')\equiv\langle \Delta(\bx)\Delta(\bx')\rangle$ gives the cosmological clustering term. Both auto-correlations contain the shot noise contribution, which is linearly proportional to the window function. The cross-correlation does not contain shot noise, since the two Poisson point processes are independent. 

In the case without IC where the density fluctuation is measured as
\begin{align}
\Delta^{{\rm wW}}(\bx)=w(\bx)\frac{n_{\rm g}(\bx)-\alpha r(\bx)}{\bar{n}^{\rm true}}\,,   
\end{align}
the above auto and cross correlations give the standard shot noise expression in the two-point function \cite{23Gebhard_SFB_eBOSS,94FKP}
\begin{align}
   &\langle \Delta^{{\rm wW},{\rm d}}(\bx)\Delta^{{\rm wW},{\rm d}}(\bx') \rangle =w(\bx)W(\bx)w(\bx')W(\bx')\xi(\bx,\bx')\nonumber\\
    &\qquad+\frac{1+\alpha}{\bar{n}^{\rm true}}w(\bx)w(\bx')W(\bx)\delta^{\rm D}(\bx-\bx')\,,
    \label{eq:shot-no-IC}
\end{align}
where we use the superscript ``d'' to indicate the discrete nature of the catalog. 

The full SFB power spectrum of the shot noise term is then \cite{21Gebhardt_SuperFab,23Gebhard_SFB_eBOSS}
\begin{align}
    (N^{\rm wW})_{\rho;}^{\phantom{*}\gamma}&=\frac{1+\alpha}{\bar{n}^{\rm true}}(w^2W)_{\rho}^{\gamma}\label{eq:sn-nic-sfb}\,,
\end{align}
and the shot noise term in the pseudo SFB power spectrum becomes
\begin{align}
N^{{\rm wW}}_{\ell n_1 n_2}&=\frac{1+\alpha}{\bar{n}^{\rm true}}\int dx\,x^2g_{n_1\ell}(x)g_{n_2\ell}(x)\nonumber\\
&\quad\frac{1}{4\pi}\int d^2\hat{\bx}\,(w^2W)(\bx)\,,
\label{eq:sn-nic-sfb-pseudo}
\end{align}
which is straightforward to compute.

\begin{figure*}[tbp]
\centerline{\includegraphics[width=0.8\textwidth]{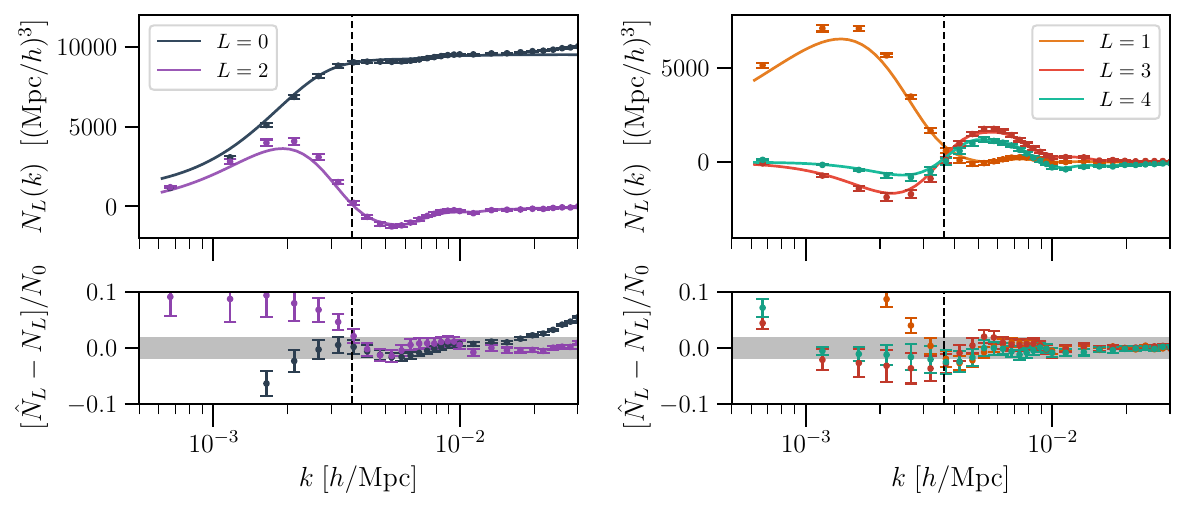}}
\caption{The theoretical and measured power spectrum multipoles $N_{L}(k)$ for the shot noise component under the realistic SPHEREx-like window with radial integral constraints. The lines show the theoretical results of the shot noise, and the points show the mean and the $1\sigma$ error of the measured Legendre multipoles from the 1,000 synthetic randoms for $z=0.2-0.5$ based on the Yamamoto estimator. Here we adopt the same plotting convention as in \cref{fig:pkl_realistic}.}
\label{fig:Nkl_realistic_RIC}
\end{figure*}

For the IC case, we rewrite the expression for measuring the density contrast given in Eq.~\eqref{eq:density-measure} as follows
\begin{align}
    \Delta^{\mathcal{O},{\rm d}}(\bx)\simeq\frac{w(\bx)}{\bar{n}^{\rm true}}\left(n_{\rm g}(\bx)-\frac{\bar{n}_{\rm g}(\bx)}{\bar{u}(\bx)}r(\bx)\right)\,,
    \label{eq:down-selection}
\end{align}
where the synthetic catalog $n_{\rm s}(\bx)$ is constructed by down-selecting the random catalog according to the ratio between the measured average densities of the galaxy and random catalogs. In practice, the estimated average number densities of the galaxy and random catalogs are obtained by averaging over some chunks of the survey volume. At the linear order, we ignore the impact of IC on the normalization factor $A$, as demonstrated in Eq.~\eqref{eq:estimate-A}. 

The auto-correlation of the galaxy catalog will have the same expression as the case without IC, while the cross-correlation between the galaxy and synthetic catalogs is
\begin{align}
\left\langle\frac{\bar{n}_{\rm g}(\bx)}{\bar{u}(\bx)}r(\bx)n_{\rm g}(\bx')\right\rangle = \langle\bar{n}_{\rm g}(\bx)n_{\rm g}(\bx')\rangle\left\langle\frac{r(\bx)}{\bar{u}(\bx)}\right\rangle\,.
\end{align}
The above product factorizes since the galaxy and random catalogs are independent. Similarly, the auto-correlation of the synthetic catalog also factorizes
\begin{align}
\left\langle \bar{n}_{\rm g}(\bx)\bar{n}_{\rm g}(\bx')\right\rangle 
 \left\langle \frac{r(\bx)}{\bar{u}(\bx)}\frac{r(\bx')}{\bar{u}(\bx')}\right\rangle \,. \nonumber
\end{align}
Unlike the case without IC, the synthetic catalog now becomes correlated with the galaxy catalog, since the galaxy catalog is partially incorporated into the generation of the synthetic catalog.

Since $u(\bx)$ is a uniform random, we will assume the averaging procedure defined in Eq.~\eqref{eq:measured-mean} yields results close to its true average density and can be taken out of the ensemble average:
\begin{align}
    &\left\langle\frac{r(\bx)}{\bar{u}(\bx)}\right\rangle \simeq \alpha(\bar{n}^{\rm true})^{-1}\langle r(\bx)\rangle =W(\bx)\\
    &\left\langle \frac{r(\bx)}{\bar{u}(\bx)}\frac{r(\bx')}{\bar{u}(\bx')}\right\rangle \simeq \alpha^2(\bar{n}^{\rm true})^{-2}\left\langle r(\bx)r(\bx')\right\rangle=W(\bx)W(\bx')\nonumber\\
    &\qquad+\alpha(\bar{n}^{\rm true})^{-1}W(\bx)\delta^{\rm D}(\bx-\bx')\label{eq:random-ratio2}\,.
\end{align}

Hereafter, we only keep tracking the shot noise contributions, as the cosmological terms under IC are already modeled in the main text. The shot noise generated from the cross correlation between the galaxy and synthetic catalog is
\begin{align}
&\left\langle\frac{\bar{n}_{\rm g}(\bx)}{\bar{u}(\bx)}r(\bx)n_{\rm g}(\bx')\right\rangle_{\rm sn} = W(\bx)\int_{\br} \,\epsilon(\bx,\br)\langle n_{\rm g}(\br)n_{\rm g}(\bx')\rangle_{\rm sn}\nonumber\\
&=\bar{n}_{\rm true}\int_{\br}\, G(\bx,\br)\delta^{\rm D}(\br-\bx')=\bar{n}_{\rm true} G(\bx,\bx')\,,
\label{eq:IC-SN-galaxy-synthetic}
\end{align}
where $G(\bx,\bx')$ is the IC kernel defined in Eq.~\eqref{eq:G-kernel-definition}. Combining Eqs.~\eqref{eq:Poisson-shot-noise} and \eqref{eq:random-ratio2}, the auto-correlation of the synthetic catalog becomes
\begin{widetext}
\begin{align}
&\left\langle\frac{\bar{n}_{\rm g}(\bx)}{\bar{u}(\bx)}r(\bx)\frac{\bar{n}_{\rm g}(\bx')}{\bar{u}(\bx')}r(\bx')\right\rangle_{\rm sn}=\int_{\br} \epsilon(\bx,\br)\int_{\br'}\epsilon(\bx',\br')\Bigg[\bar{n}_{\rm true}W(\bx)W(\bx')W(\br)\delta^{\rm D}(\br-\br')+\alpha\bar{n}_{\rm true}W(\br)W(\br')\nonumber\\
&\qquad +(1+\xi(\br,\br'))W(\bx)\delta^{\rm D}(\bx-\bx')+\alpha W(\br)\delta^{\rm D}(\br-\br')W(\bx)\delta^{\rm D}(\bx-\bx')\Bigg]\nonumber\\
&=\bar{n}_{\rm true}W(\bx')\int d^3\br \,G(\bx,\br) \epsilon(\bx',\br)+\alpha\bar{n}_{\rm true}W(\bx)\delta^{\rm D}(\bx-\bx')\nonumber\\
&\qquad+\alpha\bar{n}^{\rm true}\delta^{\rm D}(\bx-\bx')W(\bx)\langle\bar{\Delta}(\bx)\bar{\Delta}(\bx')\rangle+\alpha\delta^{\rm D}(\bx-\bx')\int d^3\br \,G(\bx,\br)\epsilon(\bx,\br)\label{eq:IC-SN-s-s}\,,
\end{align}
\end{widetext}
where $\epsilon(\bx,\br)$ is the averaging kernel defined in Eq.~\eqref{eq:measured-mean}. We observe that the second line in Eq.~\eqref{eq:IC-SN-s-s} represents the product of cosmological clustering with random shot noise, and the product of galaxy shot noise with random shot noise, respectively. These constitute the next-to-leading-order contributions compared to the galaxy shot noise and random shot noise terms in the first line of Eq.~\eqref{eq:IC-SN-s-s}. Moreover, both terms are further suppressed by factors proportional to $\alpha$. As such, we can neglect these subdominant contributions for modeling the large-scale PS.

Combining Eqs.~\eqref{eq:Poisson-shot-noise} and \eqref{eq:IC-SN-galaxy-synthetic} along with the first line of Eq.~\eqref{eq:IC-SN-s-s}, the full SFB power spectrum of the shot noise under a generic IC is
\begin{align}(N^{\mathcal{O}})_{\rho;}^{\phantom{*}\gamma}&\simeq\frac{1+\alpha}{\bar{n}^{\rm true}}(w^2W)_{\rho}^{\gamma}\nonumber\\
&\quad-\frac{1}{\bar{n}^{\rm true}}(wGw)_{\rho;}^{\phantom{*}\gamma}-\frac{1}{\bar{n}^{\rm true}}(wGw)_{\phantom{*}\rho}^{\gamma;}\nonumber\\
&\quad+\frac{1}{\bar{n}^{\rm true}}\sum_{\beta}(wG)_{\rho;}^{\phantom{*}\beta}(wW\epsilon)^{\gamma;}_{\phantom{*}\beta}\,,\label{eq:existing-sn-formula}
\end{align}
where $wGw(\bx,\bx')\equiv w(\bx)G(\bx,\bx')w(\bx')$, and $wW\epsilon\equiv w(\bx)W(\bx)\epsilon(\bx,\bx')$.  The above expressions are obtained by performing the bivariate SFB decomposition (\cref{eq:bivariate-inverse}) to all the kernel functions and applying the orthogonality of the dSFB basis. 

After this rigorous derivation, we note that Eq.~\eqref{eq:existing-sn-formula} can alternatively be obtained via a simplified—though less rigorous—argument, analogous to the cosmological clustering term in Eq.~\eqref{eq:SFB-full-PS-IC}:
\begin{align}
(N^{\mathcal{O}})_{\rho;}^{\phantom{*}\gamma}\simeq\sum_{\mu,\nu}(wW-wG)_{\rho;}^{\phantom{x}\mu}(wW-wG)^{\gamma;}_{\phantom{x}\nu} N_{\mu;}^{\phantom{*}\nu}\,,\label{eq:sn-simple}
\end{align}
where the theoretical shot noise is defined as
\begin{align}
   N_{\mu;}^{\phantom{*}\nu}\equiv\frac{1}{\bar{n}^{\rm true}} (W_{\nu}^{\mu})^{-1}\label{eq:theory-shot-noise}\,.
\end{align}
While this definition of the theoretical shot noise is not formally well-defined—since any realistic window function $W(\bx)$ lacks a true inverse due to incomplete sky coverage—it nonetheless serves as a useful analytic shorthand. In particular, substituting Eq.~\eqref{eq:theory-shot-noise} into Eq.~\eqref{eq:sn-simple} directly yields Eq.~\eqref{eq:existing-sn-formula} in a straightforward manner. This was the approach taken in Ref.~\cite{23Gebhard_SFB_eBOSS}. 

The above heuristic argument can be made more formal by noting that the window function is invertible in the idealized case of full-sky coverage, even when different regions are observed with varying depth. One may then introduce a small parameter~$\epsilon$ such that regions with no coverage are assigned values of order~$\epsilon$, rendering the window function invertible and the theoretical shot noise term well-defined for any finite~$\epsilon$. The substitution of Eq.~\eqref{eq:theory-shot-noise} into Eq.~\eqref{eq:sn-simple} can therefore be carried out formally for finite~$\epsilon$. The resulting expression, Eq.~\eqref{eq:existing-sn-formula}, will not contain the explicit inverse of the window. Consequently, the limit $\epsilon\to 0$ can be safely taken, and this regularization procedure is well defined.

For the global IC case, the shot noise expression will simplify to
\begin{align}(N^{\mathcal{O}})_{\rho;}^{\phantom{*}\gamma}&\simeq\frac{1+\alpha}{\bar{n}^{\rm true}}(w^2W)_{\rho}^{\gamma}\nonumber\\
&\quad-\frac{1}{\bar{n}^{\rm true}I_{10}}(wW)_{\rho}(wW)^{\gamma}\label{eq:sn-gic}\,.
\end{align}
With Eq.~\eqref{eq:GIC-SFB}, one can easily show the second and third terms $wGw$ of Eq.~\eqref{eq:existing-sn-formula} are the same as the second line in the above Eq.~\eqref{eq:sn-gic}. For the fourth term, note that
\begin{align}
(wW\epsilon)^{\gamma;}_{\phantom{*}\beta}
&=\frac{1}{I_{10}}(wW)^{\gamma}d_{\beta}\,,
\label{eq:wWepsilon-GIC}
\end{align}
where $d_\beta$ is defined in Eq.~\eqref{eq:dmu}. Applying Eq.~\eqref{eq:dnlm}, we get
\begin{align}
\sum_{\beta}(wG)_{\rho;}^{\phantom{*}\beta}(wW\epsilon)^{\gamma;}_{\phantom{*}\beta}&=\sum_{\beta} \frac{1}{I_{10}}(wW)_{\rho}W^{\beta}\frac{1}{I_{10}}(wW)^{\gamma}d_{\beta}\nonumber\\
&=\frac{1}{I_{10}}(wW)_{\rho}(wW)^{\gamma}\,.
\end{align}
Therefore, the second, third, and fourth terms in Eq.~\eqref{eq:existing-sn-formula} are the same under the global IC, leaving only one IC term in the final shot noise expression after cancellation.

The radial IC case has a similar structure where the second, third, and fourth terms in Eq.~\eqref{eq:existing-sn-formula} are also the same, resulting in 
\begin{align}
(N^{\mathcal{O}})_{\rho;}^{\phantom{*}\gamma}&\simeq\frac{1+\alpha}{\bar{n}^{\rm true}}(w^2W)_{\rho}^{\gamma}\nonumber\\
&\quad-\frac{4\pi}{\bar{n}^{\rm true}}\sum_{n'}(wW)_{n_\rho\ell_\rho m_\rho}^{n'00}(w\widetilde{W})^{n_{\gamma}\ell_{\gamma}m_{\gamma}}_{n'00}\label{eq:sn-RIC}\,.
\end{align}
The second and third terms in Eq.~\eqref{eq:existing-sn-formula} are the same since the radial IC kernel $G(\bx,\br)$ in Eq.~\eqref{eq:RIC-Gkernel} is symmetric between the two variables, and they can be easily expressed using Eq.~\eqref{eq:RIC-SFB}. For the fourth term, we note that
\begin{align}
(wW\epsilon)^{\gamma;}_{\phantom{*}\beta}&=
     \int_\bx\, \frac{1}{x^2}\delta^{\rm D}(x-r)e^{\gamma}(\bx) (w\widetilde{W})(\bx)\int_\br\, e_{\beta}(\br)\nonumber\\
&=4\pi\sum_{n''}(w\widetilde{W})^{n_\gamma\ell_\gamma m_\gamma}_{n''00}d_{n_{\beta}\ell_{\beta}m_{\beta}}^{n''00}\nonumber\\
&=4\pi\delta_{\ell_{\beta}0}^{\rm K}\delta_{m_{\beta}0}^{\rm K}(w\widetilde{W})^{n_\gamma\ell_\gamma m_\gamma}_{n_\beta 00}\,.
\label{eq:wWepsilon-RIC}
\end{align}
With $\int_{\hat{\bx}} \widetilde{W}(\bx)=1$, we have $\widetilde{W}^{n00}_{n'00}=(4\pi)^{-1}\delta_{n n'}^{\rm K}$. Applying these results, the fourth term becomes
\begin{align}
&\quad(4\pi)^2\sum_{n',n_\beta}(wW)_{n_\rho\ell_\rho m_\rho}^{n'00}\widetilde{W}^{n_{\beta}00}_{n'00}(w\widetilde{W})^{n_\gamma\ell_\gamma m_\gamma}_{n_\beta 00}\nonumber\\
&=4\pi\sum_{n'}(wW)_{n_\rho\ell_\rho m_\rho}^{n'00}(w\widetilde{W})^{n_\gamma\ell_\gamma m_\gamma}_{n' 00}\label{eq:sn-4-RIC}\,,
\end{align}
which indeed cancels with either the second or the third term.

With the full dSFB PS expressions in Eqs.~\eqref{eq:sn-gic} and \eqref{eq:sn-RIC}, one can derive the corresponding shot noise contributions to the pseudo dSFB PS. These can then be mapped to the Fourier-space PSM via Eq.~\eqref{eq:PSM-to-dSFB}. We show the shot noise PSM under the SPHEREx-like window with the radial IC applied in \cref{fig:Nkl_realistic_RIC}. At small scales, only the monopole exhibits a constant shot noise power, while all higher-order Legendre multipoles vanish—as expected for an isotropic noise. At the largest scales, however, all Legendre multipoles acquire nonzero power due to RIC, with shapes resembling the cosmological signal under RIC shown in \cref{fig:pkl_realistic}. Overall, the theoretical predictions for the shot noise PSM show reasonable agreement with the mean of the Yamamoto measurements from the synthetic randoms.

\subsection{Consistency with Ref.~\cite{23Gebhard_SFB_eBOSS}}\label{sec:consistency}

For the radial IC, referred to as the radial local average effect in Ref.~\cite{23Gebhard_SFB_eBOSS}, we note that their Eq.~(C62)  has the same form as the second and third terms $\mathcal{M}[wW,wG]+\mathcal{M}[wG,wW]$ defined in our Eq.~\eqref{eq:A-mixing}, with $wG$ given in Eq.~\eqref{eq:RIC-SFB}. With some care, one can also recognize that their Eq.~(C50), where the azimuthally averaged contraction operator $\mathcal{M}$ was applied separately to $wW$ and $\widetilde{W}$, is equivalent to the fourth term $\mathcal{M}[wG,wG]$ in this work. Our shot noise result Eq.~\eqref{eq:sn-RIC} agrees with their Eq.~(C57) for the full SFB power spectrum and their Eq.~(C46) for the pseudo SFB PS.

Ref.~\cite{23Gebhard_SFB_eBOSS} describes the global IC as the constant average density contrast case for the local average effect. In this case, it is straightforward to see that the mixing matrix $\widetilde{T}$ defined in their Eq.~(C31) to be equivalent to the terms $\mathcal{M}[wW,wG]+\mathcal{M}[wG,wW]$ in Eq.~\eqref{eq:A-mixing}, with $wG$ specified in Eq.~\eqref{eq:GIC-SFB}. 

For the fourth term of the full SFB PS, Eq.~(C33) of Ref.~\cite{23Gebhard_SFB_eBOSS} gives (we use the notations defined in \cref{sec:notation-convolution} to present the previous result):
\begin{align}
\langle (\bar{\Delta}^{wW})_{\rho}(\bar{\Delta}^{wW})^{\gamma}\rangle
&=\sum_{\epsilon,\mu,\nu,\lambda}(wW)_{\rho}^{\epsilon}D_{\epsilon}^{\mu} \nonumber\\
&\quad\langle(\Delta^{W})_{\mu}(\Delta^{W})^{\nu}\rangle
D_{\nu}^{\lambda} (wW)_{\lambda} ^{\gamma}\,,\label{eq:GIC-4thterm-old}
\end{align}
where $D_{\epsilon}^{\mu} \equiv d_\epsilon \, d^\mu$, and $d_\epsilon$ is defined in Eq.~\eqref{eq:dmu}. 

To show the equivalence, we need the following identity that relates a double SFB transform of $W(\bx)$ to a single transform:
\begin{align}
&\sum_{\beta}W_{\epsilon}^{\beta}d_{\beta}=\sum_{n_{\rho}}W_{\epsilon}^{n_{\beta}00}d_{n_\beta 00}=\int d^3\bx\, e_{\epsilon}(\bx)W(\bx)\nonumber\\
&\qquad\int dr\,r^2\sum_{n_\beta}g_{n_{\beta}0}(x)g_{n_{\beta}0}(r)=W_{\epsilon}\,,
\end{align}
where we applied Eqs.~\eqref{eq:dmu} and \eqref{eq:dSB-complete}. 

Applying the above identity and rearranging terms, Eq.~\eqref{eq:GIC-4thterm-old} becomes
\begin{align}
\langle (\bar{\Delta}^{wW})_{\rho}(\bar{\Delta}^{wW})^{\gamma}\rangle
=\sum_{\mu,\nu}(wW)_{\rho}W^{\mu}\langle\Delta_{\mu}\Delta^{\nu}\rangle
W_{\nu}(wW)^{\gamma}\nonumber\,,
\end{align}
which agrees with our results in Eqs.~\eqref{eq:SFB-full-PS-IC} and \eqref{eq:GIC-SFB}.

\section{Review on Plane-parallel PSM}\label{sec:standard}

\subsection{Window function}\label{sec:Cartesian-window}
We here summarize the derivation and results presented in Refs.~\cite{19BeutlerMutipolePk,21BeutlerWAWindow,22CatorinaGR-P} for modeling the effects of window functions in the PSM measured from the Yamamoto estimator. These window convolution results through Cartesian coordinates have become standard in clustering analyses \cite{14BeutlerBOSS,17Beutler_BOSS,24DESI_II,25Elkhashab}.

The two-point correlation function and its multipoles are defined as
\begin{align}  \langle\Delta(\bx_1)\Delta(\bx_2)\rangle=\xi(\bs,\bx_1)=\sum_{L}\xi_{L}(s,x_1)\mathcal{L}_{L}(\hat{\bs}\cdot\hat{\bx}_1)\,,
\label{eq:2CFM}
\end{align}
where $\bs\equiv\bx_1-\bx_2$ is the separation between two galaxies.

Changing integration variables from $(\bx_1,\bx_2)$ to $(\bx_1,\bs)$
and applying Eqs.~\eqref{eq:2CFM}, \eqref{eq:integrate_L_ell}, and \eqref{eq:contraction_legendre}, the PSM of Eq.~\eqref{eq:Pl_average} under window function can be exactly expressed as \cite{22CatorinaGR-P}
\begin{widetext}
\begin{align}
P_L^{\rm W}(k)&=\frac{2L+1}{I_{22}}i^{-L}  \int ds\, s^2j_L(ks) \sum_{L'}\int dx_1\, x_1^2 \xi_{L'}(s,x_1) \int d^2\hat{\bs}\int  d^2\hat{\bx}_1\,\mathcal{L}_{L'}(\hat{\bx}_1\cdot\hat{\bs})\mathcal{L}_{L}(\hat{\bx}_1\cdot\hat{\bs})W(\bx_1)W(\bx_1-\bs)\nonumber\\
&=\frac{2L+1}{I_{22}}i^{-L} \sum_{L',L_1} \begin{pmatrix}
L & L' & L_1\\
0 & 0 & 0
\end{pmatrix}^2(2L_1+1)\int ds\, s^2j_L(ks) \int dx_1\, x_1^2 \xi_{L'}(s,x_1) F_{L_1}(s,x_1)\,,
\label{eq:PL-window-standard-exact}
\end{align}   
\end{widetext}
where we introduced
\begin{align}
    F_{L_1}(s,x_1)\equiv\int_{\hat{\bs},\hat{\bx}_1}\,\mathcal{L}_{L_1}(\hat{\bx}_1\cdot\hat{\bs})W(\bx_1)W(\bx_1-\bs)\,.
\label{eq:Fl-sx1}
\end{align}
In practice, it is computationally difficult to measure the 2D function $F_{L_1}(s,x_1)$ for realistic surveys and perform the 2D integration in Eq.~\eqref{eq:PL-window-standard-exact}, so further simplifications are required.

If the 2CFM can be evaluated at some effective redshift $z_{\rm eff}$ for the redshift bin (see next subsection for discussion), we can take 2CFM outside the integration of $x_1$ and then integrate $F_{L_1}(s,x_1)$ along $x_1$ \cite{22CatorinaGR-P,19BeutlerMutipolePk}
\begin{align}
P^{{\rm W, eff}}_L(k)&=\frac{2L+1}{I_{22}}(4\pi)i^{-L} \sum_{L,L_1} \begin{pmatrix}
L & L' & L_1\\
0 & 0 & 0
\end{pmatrix}^2\nonumber\\
&\quad\int ds\, s^2j_L(ks)\xi_{L'}(s,z_{\rm eff})Q_{L_1}(s),
\label{eq:PL-window-eff-s}
\end{align}
where we defined the configuration-space window function multipole
\begin{align}
Q_{L_1}(s)&\equiv\frac{2L_1+1}{4\pi}\int d^2\hat{\bs}\int d^3\bx_1 \mathcal{L}_{L_1}(\hat{\bx}_1\cdot\hat{\bs})\nonumber\\
&\qquad W(\bx_1)W(\bx_1-\bs)\label{eq:QLs}\,.
\end{align}
Eq.~\eqref{eq:PL-window-eff-s} models the window convolution of PSM through Cartesian configuration space, which was adopted in earlier works such as Refs.~\cite{17Wilson,17Beutler_BOSS,19BeutlerMutipolePk}. One can obtain the window function multipole of Eq.~\eqref{eq:QLs} either through pair counting \cite{17Wilson} or a combination of Hankel transform and FFT \cite{19BeutlerMutipolePk}.

To avoid going through the configuration space for window function modeling, we can further express the 2CFM as the Hankel transform of unwindowed PSM
\begin{equation}
\xi_{L}(s,x_1)=\frac{i^L}{2\pi^2}\int dk\,k^2 P^{\rm loc}_{L}(k,x_1)j_{L}(ks)\label{eq:XiL-PkL}\,,
\end{equation}
where the unwindowed, local power spectrum is defined as the Fourier transform of the 2CF over the galaxy separation $\bs$ \cite{15Scoccimarro_FKP,23WAGR}
\begin{align}
P^{\rm loc}(\bk,\bx_1)&\equiv\int \dd^3\bs\; e^{-i\bk\cdot\bs}\xi(\bs,\bx_1)\nonumber\\
&=P^{\rm loc}(k,\mu=\hat{\bk}\cdot\hat{\bx}_1,x_1)\label{eq:local-PS}\\
&=\sum_{L}P_{L}^{\rm loc}(k,x_1)\mathcal{L}
(\hat{\bk}\cdot\hat{\bx}_1)\label{eq:local-PSM}\,.
\end{align}
Using Eq.~\eqref{eq:integrate_L_ell}, one can verify that the unwindowed, local PSM and 2CFM indeed form a Hankel transform pair.

Using Eq.~\eqref{eq:XiL-PkL}, one can rewrite Eq.~\eqref{eq:PL-window-eff-s} in Cartesian Fourier space \cite{14BeutlerBOSS,21BeutlerWAWindow} 
\begin{align}
P^{\rm W,eff}_L(k)&=\int\,dk'\, \sum_{L'}k'^2  W_{LL'}(k,k') P^{\rm loc}_{L'}(k',z_{\rm eff})\,,
\label{eq:PL-window-eff-k}
\end{align}
where the Fourier space window mixing matrix can be obtained by a double spherical Bessel integral of the window function multipole $Q_{L}(s)$
\begin{align}
W_{LL'}(k,k')&=\frac{2}{\pi}\frac{2L+1}{I_{22}} i^{L'-L}  \sum_{L_1}\begin{pmatrix}
L & L' & L_1\\
0 & 0 & 0
\end{pmatrix}^2 \nonumber\\
&\quad \int ds\, s^{2} j_L(ks) j_{L'}(k's)  Q_{L_1}(s)\,.
\label{eq:WLLpkkp}
\end{align}
Eq.~\eqref{eq:PL-window-eff-k} expresses the windowed PSM, as estimated by the Yamamoto estimator, as a matrix-weighted sum over the unwindowed, local PSM. This approach has been adopted in most of the recent works, such as Refs.~\cite{21BeutlerWAWindow,24Cagliari_eBOSS_quasar_fnl,24DESI_II,25Elkhashab}. In practice, the mixing matrix Eq.~\eqref{eq:WLLpkkp} can be computed from \texttt{pypower} \cite{24DESI_II}.

We emphasize that the Cartesian Fourier space approach of Eq.~\eqref{eq:PL-window-eff-k} relies on the effective redshift approximation. An exact treatment in Cartesian coordinates instead requires evaluating the computationally challenging expression in Eq.~\eqref{eq:PL-window-standard-exact}. Even in the simplified case considered in \cref{sec:WA} where redshift evolution is neglected (corresponding to a snapshot of simulations), that is $b_1$, $f$ and $D$ are constants, the local PSM and 2CFM still depend on the galaxy pair distance $x_1$ at large scales due to wide-angle effects (see Eq.~(5.6) of Ref.~\cite{22CatorinaGR-P}). Therefore, Eq.~\eqref{eq:PL-window-eff-k} remains inexact on large scales, even when redshift evolution is absent. 

However, at small scales where the plane-parallel approximation holds, the Kaiser formulas in Eqs.~\eqref{eq:kaiser} and \eqref{eq:kaiser-coeff} below yield local PSM and 2CFM that are independent of the galaxy pair position $\bx_1$ in the absence of redshift evolution. In this regime, Eq.~\eqref{eq:PL-window-eff-k} becomes exact. Consequently, the Cartesian Fourier-space approach of Eq.~\eqref{eq:PL-window-eff-k} and the dSFB-space approach of Eq.~\eqref{eq:IC-model-SFB} are expected to agree at small scales—which is indeed confirmed by the results shown in \cref{fig:WA_even}.

\subsection{Effective redshift approximation}\label{sec:ERA}

Due to the computational challenge of exactly modeling the effects of window functions in Cartesian coordinates as discussed above, the effective redshift approximation has been adopted in most 3D clustering analyses using 2CFM and PSM to date. Under this approximation, one only needs to compare the clustering measurement with the model computed at a single effective redshift, rendering the calculation more tractable.

One can define a common effective redshift for the redshift bin shared by all multipoles \cite{19Castorina_fnl_quasar,24Cagliari_eBOSS_quasar_fnl} 
\begin{align}
z_{\rm eff}\equiv\frac{\int d^3\bx\,w(\bx)^2W(\bx)^2 z(x)}{\int d^3\bx\,w(\bx)^2W(\bx)^2}\,
\label{eq:global_zeff}
\end{align}
Eq.~\eqref{eq:global_zeff} is defined in the power of $w(\bx)^2W(\bx)^2$, and Ref.~\cite{24Cagliari_eBOSS_quasar_fnl} has shown this definition of the common effective redshift to be more accurate than the other commonly used version \cite{20Neveux_eBOSS_quasar_PS,21Hou_eBOSS_quasar_CF,22Mueller_eBOSS_quasar_fnl} that is linearly proportional to $w(\bx)W(\bx)$, so we adopt Eq.~\eqref{eq:global_zeff}.

Due to the anisotropies of galaxy clustering caused by RSD, it is more accurate to define an effective redshift separately for each multipole \cite{22CatorinaGR-P}\footnote{Ref.~\cite{22CatorinaGR-P} defines $z_{{\rm eff},L}$ in terms of the average of the 2CFM, which is mathematically equivalent to the definition in Eq.~\eqref{eq:local_zeff}, since 2CFM and PSM form a Hankel transform pair.}
\begin{align}
    P^{\rm loc}_{L}(k,z_{{\rm eff},L})\equiv \frac{\int d^3\bx \,w(\bx)^2W(\bx)^2P^{\rm loc}_{L}(k,z(x))}{\int d^3\bx\,w(\bx)^2W(\bx)^2}\,,
    \label{eq:local_zeff}
\end{align}
where the multipole-dependent effective redshift $z_{{\rm eff},L}$ is defined through the weighted volume average of the local PSM. The common version is significantly easier to evaluate and only needs to be computed once (assuming a fixed redshift-to-distance relation), while the multipole version depends on the cosmological modeling of the power spectrum. The integral in Eq.~\eqref{eq:local_zeff} is more difficult to compute, explaining the wide adoption of the common ERA. 

Assuming linear theory and the p.p. approximation significantly simplify the evaluation of Eq.~\eqref{eq:local_zeff}. In the Kaiser formula, the redshift (time) dependence of the PSM can be factored out as a coefficient independent of the Fourier modes \cite{87Kaiser}
\begin{align}
    P^{\rm loc}_{L}(k,z)&=p_{L}(z)P_{m,0}(k)\label{eq:kaiser},
\end{align}
where $P_{m,0}(z)$ denotes the matter power spectrum at the present time, and the redshift evolution coefficient $p_{L}(z)$ is given by
\begin{subequations}
\begin{align}
    p_0&=\left(b_1^2+\frac{2}{3}b_1f+\frac{1}{5}f^2\right)D^2\\
    p_2&=\left(\frac{4}{3} b_1 f+\frac{4}{7} f^2\right)D^2\\
    p_4&=\frac{8}{35}f^2D^2\,,
\end{align}
\label{eq:kaiser-coeff}
\end{subequations}
where $b_1$, $D$ and $f$ all depend on the redshift. Therefore, the multipole ERA defined in Eq.~\eqref{eq:local_zeff} simplifies to \begin{align}
    p_{L}(z_{{\rm eff},L})=\frac{\int d^3\bx\, w(\bx)^2W(\bx)^2 p_{L}(z)}{\int d^3\bx\, w(\bx)^2W(\bx)^2}\,.\label{eq:zeff_l}
\end{align}
One can then calculate the multipole-dependent effective redshift with
\begin{align}
    z_{{\rm eff},L}=p_{L}^{-1}\left(\frac{\int d^3\bx\, w(\bx)^2W(\bx)^2 p_{L}(z)}{\int d^3\bx\, w(\bx)^2W(\bx)^2}\right)\,,
\end{align}
where $p_L^{-1}$ is the inverse function of the the redshift evolution coefficient. We used the above formula for multipole ERA in \cref{sec:RE}.

\section{Mapping Extensions}\label{sec:detail_gsfb_to_psm}
\subsection{Derivative formulation}\label{sec:deriv-derivation}
We present here a new formulation, which is based on the derivatives of spherical Bessel functions, for the mapping between the gSFB power spectrum and the PSM given in Eq.~\eqref{eq:PSM-gSFB}. This approach differs from the earlier derivations presented in Refs.~\cite{24PSM_SFB,18CastorinaWA}.

Starting from Eq.~\eqref{eq:fl_estimator}, the Fourier transform of the field weighted by a Legendre polynomial, we have
\begin{align}
F_L(\bk)&=\int_{\bx}e^{-i\bk\cdot\bx} \sum_{n=0}^{L}c_{Ln}(\hat{\bk}\cdot\hat{\bx})^n\Delta(\bx)\nonumber\\
&=\sum_{n=0}^{L}c_{Ln}i^{n}\int_{\bx}\Delta(\bx)\left[\frac{\partial^n}{\partial (kx)^n}e^{-i\bk\cdot\bx}\right]\nonumber\\
&=(4\pi)\sum_{n=0}^{L}c_{Ln}\sum_{\ell,m}i^{n-\ell}Y_{\ell m}(\hat{\bk})\Delta_{\ell m}^{(n)}(k)
\label{eq:FLk-derivative-SFB}
\end{align}
where we first directly expressed the Legendre polynomials as polynomials $\mathcal{L}_L(x)=\sum_{n=0}^{L}c_{Ln}x^n$, applied the Rayleigh wave expansion (\cref{eq:plane_wave_exp-Y}), and then defined the following $n$-th derivative SFB mode
\begin{align}
\Delta_{\ell m}^{(n)}(k) &=\int_{\bx}j^{(n)}_{\ell}(kx)Y^*_{\ell m}(\hat{\bx})\Delta(\bx)\,.
\label{eq:derivative-sfb_x-to-k}
\end{align}
Compared to the standard SFB mode of Eq.~\eqref{eq:sfb_x-to-k}, we now use the $n$-th order derivative of spherical Bessel functions in the radial transform. We note that Eq.~\eqref{eq:FLk-derivative-SFB} generalizes Eq.~\eqref{eq:deltaklm2deltak} to higher multipoles $L>0$ such that the Legendre-weighted Cartesian Fourier mode can be expressed as a sum of the derivative SFB modes.

One can then form the derivative SFB power spectrum
\begin{align}
   \langle\Delta_{\ell_1 m_1}^{(n)}(k_1) \Delta_{\ell_2 m_2}^{(n')*}(k_2)\rangle = C_{\ell_1}^{(n,n')}(k_1,k_2)\delta_{\ell_1\ell_2}^{\rm K}\delta_{m_1m_2}^{\rm K}\,.
   \label{eq:derivative-SFBPS-def}
\end{align} 
Applying Eq.~\eqref{eq:YY_ortho}, we can express the PSM from Yamamoto estimator (\cref{eq:pl_estimator}) as
\begin{align}
&\frac{I_{22}}{4\pi(2L+1)}P^{\mathcal{O}}_L(k)\equiv\int d^2\hat{\bk}\sum_{n=0}^{L}c_{Ln}\sum_{b,m_b}i^{n-b}Y_{b m_b}(\hat{\bk})\nonumber\\
&\qquad\qquad\sum_{a,m_a}i^{a}Y_{a m_a}^*(\hat{\bk})\langle\Delta_{bm_b}^{(n),\mathcal{O}}(k)\Delta_{a m_a}^{\mathcal{O}*}(k)\rangle\nonumber\\
&\qquad=\sum_{b}(2b+1)\left[\sum_{n=0}^{L}c_{Ln}i^{n}C_{b}^{(n,0),\mathcal{O}}(k,k)\right]\,,
\label{eq:PSM-derivativeSFB}
\end{align}
which is the sum of the derivative SFB power spectra weighted by the coefficients of Legendre polynomials. 

For example, the PS dipole and quadrupole can be expressed as
\begin{align}
P_{1}(k)&=i\frac{12\pi}{I_{22}}\sum_{b}(2b+1)C_{b}^{(1,0)}(k,k) \\
P_{2}(k)&=-\frac{20\pi}{I_{22}}\sum_{b}(2b+1)\Bigg(\frac{1}{2}C_{b}(k,k)\nonumber\\
&\qquad+\frac{3}{2}C_{b}^{(2,0)}(k,k)\Bigg)\,.
\end{align}
The PS multipole of order $L$ incorporates contributions from the derivative SFB PS up to the $L^{\rm th}$-order. While these derivative SFB PS are evaluated on the diagonal $k_1=k_2=k$, the presence of spherical Bessel function derivatives means that the $L>0$ multipoles effectively encode off-diagonal information from the SFB PS. In particular, higher-order PSM multipoles capture increasingly fine-grained information on RSD and redshift evolution due to the use of higher-order radial derivatives. We note that the derivatives contained in the SFB PS are not symmetric, a feature arising from the end-point line of sight adopted in the Yamamoto estimator.

Since the derivative formulation of Eq.~\eqref{eq:PSM-derivativeSFB} has to be equivalent to the gSFB-to-PSM mapping of Eq.~\eqref{eq:PSM-gSFB}, the following identity needs to hold
\begin{align}
&\quad\sum_{a}i^{-a+b}(2a+1)\begin{pmatrix}
a & L & b\\
0 & 0 & 0
\end{pmatrix}^2 j_a(kx)\nonumber\\
&=\sum_{n=0}^{L}c_{Ln}i^{n}j^{(n)}_b(kx)\,,
\label{eq:Legendre_SBderiv-Wigner_SB}
\end{align}
where the sum over spherical Bessel functions involving Wigner-3$j$ equals to the sum over higher-order derivatives of spherical Bessel functions weighted by the coefficients of Legendre polynomials. We will directly prove the above identity in Eq.~\eqref{eq:Flbkx}.

\subsection{Mapping with cSFB}
Similar to Eq.~\eqref{eq:d-to-g} of \cref{sec:dSFB-gSFB} where we expressed the generalized SFB mode to a sum of the discrete SFB modes, we can also express the gSFB mode as a convolution of the continuous SFB mode:
\begin{align}
\Delta^{a}_{\ell m}(k)=\frac{2}{\pi}\int dq\, q^2\mathcal{I}_{\ell a}(k,q)\Delta_{\ell m}(q)\,,
\end{align}
where $\mathcal{I}_{\ell \ell'}(k,q)$ is the integral over two spherical Bessel functions of different orders
\begin{align}
\mathcal{I}_{\ell \ell'}(k,q)\equiv\int_{0}^{\infty} dr\, r^2 j_{\ell}(kr)j_{\ell'}(qr)\,.
\label{eq:double-Bessel}
\end{align}

The results for the above integral have been derived in Refs.~\cite{91Maximon_2SB,19Mehrem_Gaussian,23PaulWA}, and applied in a cosmological context in Ref.~\cite{23PaulWA}. The resulting analytical expressions involve the Dirac delta functions, Heaviside step functions, and hypergeometric functions. While exact, these expressions pose significant challenges for further analytical integration and are often numerically unstable due to the highly peaked nature of the involved functions. In contrast, the dSFB-to-gSFB mapping functions exhibit much greater numerical stability, which is one of the reasons for our preference for the dSFB basis. 

For completeness, we can express the gSFB PS as a 2D-convolution of the cSFB PS
\begin{align}
C_{\ell}^{ab}(k_1,k_2)&=\frac{4}{\pi^2}\int dq\,q^2\mathcal{I}_{\ell a}(k_1,q)\nonumber\\
&\quad\int dq'\,q'^2\mathcal{I}_{\ell b}(k_2,q')C_{\ell}(q,q')\,,
\end{align}
and we can rewrite the gSFB-to-PSM mapping of Eq.~\eqref{eq:PSM-gSFB} in terms of the cSFB PS
\begin{align}
P^{\mathcal{O}}_{L}(k)&=\frac{4\pi(2L+1)}{I_{22}}\sum_{b}(2b+1)\nonumber\\
&\quad\int dq\,q^2\mathcal{T}_{b}^{L}(k,q)C^{\mathcal{O}}_{b}(q,k)\,,
\label{eq:cSFB-to-PSM}
\end{align}
where we define the following mapping function
\begin{align}
\mathcal{T}_{b}^{L}(k,q)=\frac{2}{\pi}\sum_{a}i^{-a+b}(2a+1)\begin{pmatrix}
a & L & b\\
0 & 0 & 0
\end{pmatrix}^2\mathcal{I}_{ba}(k,q)\nonumber\,.
\end{align}
Equation~\eqref{eq:cSFB-to-PSM} explicitly demonstrates that PSM with $L>0$ contains off-diagonal components of the cSFB PS. Both the generalized SFB and derivative SFB modes serve as convenient intermediate quantities that allow the incorporation of the off-diagonal information, while still evaluating the SFB power spectrum on the diagonal $k_1=k_2=k$ to exactly match the magnitude of the Fourier mode $k$ used in PSM.

\subsection{Alternative LOS}\label{sec:other-LOS}

The end-point LOS is the default choice in the Yamamoto estimator due to its factorization property, as seen in Eq.~\eqref{eq:pl_estimator}, which enables fast and efficient implementations using FFTs. All large-scale structure analyses have adopted this convention to date. Alternative LOS definitions, such as the mid-point and angular bisector, have also been considered in literature:
\begin{align}
    \hat{\bx}_{\rm bisector}&=\frac{\hat{\bx}_1+\hat{\bx}_2}{|\hat{\bx}_1+\hat{\bx}_2|}\,,\label{eq:bisector}\\
    \hat{\bx}_{\rm mid}&=\frac{\bx_1+\bx_2}{|\bx_1+\bx_2|}\label{eq:mid}\,,
\end{align}
where $\bx_1$ and $\bx_2$ are the positions for the two galaxies in a pair. These definitions enforce symmetry between the galaxy pair and lead to a suppression of wide-angle corrections  compared to the end-point LOS\footnote{Despite this suppression, wide-angle effects are present for any LOS definition and can be modeled exactly on the theory side.}\cite{18CastorinaWA,19BeutlerMutipolePk}.

However, estimators built on mid-point or angular bisector LOS are computationally challenging, as they break the pairwise factorization and require explicit summation over all galaxy pairs, which becomes prohibitively expensive for Stage-IV surveys. Ref.~\cite{21Philcox_bimid_estimator} introduced fast, factorizable implementations for mid-point and bisector LOS using FFTs by expanding in a perturbative series over the galaxy pair separation, analogous to perturbative wide-angle modeling on the theory side. Nevertheless, this expansion does break down for galaxy pairs with large angular separations. Moreover, both mid-point and bisector line of sights can become ill-defined for specific configurations—namely when $\hat{\bx}_1+\hat{\bx}_2=0$ or $\bx_1+\bx_2=0$ in wide-field surveys. Due to these challenges in both efficiency and robustness for alternative choices, we expect the end-point LOS to remain the default in practical analyses.

That said, in certain situations, alternative LOS definitions can potentially offer advantages. In particular, the unwindowed PS (or 2CF) dipole has been recognized as a probe of relativistic effects, especially the Doppler term \cite{14Bonvin_asymmetry,23Bonvin_cross}. Under the end-point LOS, this signal is significantly contaminated by wide-angle leakage from the monopole into the dipole. In contrast, the mid-point and bisector LOS choices suppress such contamination, making them better suited for isolating general relativistic contributions from the dominant Newtonian RSD terms\footnote{For the purpose of constraining fundamental physics, it is not strictly necessary to isolate relativistic effects from the Newtonian RSD terms, since one can consistently model all contributions to obtain parameter constraints.}. This potential benefit motivates our discussion of the SFB-to-PSM mapping under a more general choice of LOS.

Under a generic LOS $\hat{\bx}_{\rm c}(\bx_1,\bx_2)$, that can be an arbitrary function of the positions of both galaxies in a pair, the PSM becomes
\begin{align}
P^{\mathcal{O}}_L (k)&= \frac{2L+1}{I_{22}} \int\frac{d^2\hat{\bk}}{4\pi}\int_{\bx_1,\bx_2} e^{-i \bk \cdot (\bx_1-\bx_2)} \nonumber\\
&\qquad \langle \Delta^{\mathcal{O}}(\bx_1)\Delta^{\mathcal{O}}(\bx_2) \rangle \mathcal{L}_L(\hat{\bx}_{\rm c}\cdot \hat{\bk})\,,
\label{eq:PL_generic_LOS}
\end{align}
Applying the Rayleigh wave expansion (\cref{eq:plane_wave_exp-L}) on the exponential and the dSFB decomposition (\cref{eq:dsfb_x-to-k}) on the field, we obtain

\begin{widetext}
\begin{align}
P^{\mathcal{O}}_{L}(k)&=\sum_{n_1,\ell_1,m_1}\sum_{n_2,\ell_2,m_2}\sum_{a,b}\mathcal{Q}_{n_1\ell_1m_1,n_2\ell_2m_2}^{L,ab}(k)\langle\Delta^{\mathcal{O}}_{n_1\ell_1m_1}\Delta^{*\mathcal{O}}_{n_2\ell_2m_2}\rangle\,,
\label{eq:generic-dSFB-to-PSM}
\end{align}
where the coupling matrix is
\begin{align}
    \mathcal{Q}_{n_1\ell_1m_1,n_2\ell_2m_2}^{L,ab}(k)&\equiv\frac{(2L+1)}{4\pi I_{22}}i^{-a+b}(2a+1)(2b+1)\int_{\bx_1}j_a(kx_1)g_{n_1\ell_1}(x_1)Y_{\ell_1m_1}(\hat{\bx}_1)\int_{\bx_2}j_b(kx_2)g_{n_2\ell_2}(x_2)Y^{*}_{\ell_2m_2}(\hat{\bx}_2)\nonumber\\
    &\int_{\hat{\bk}}\mathcal{L}_{a}(\hat{\bk}\cdot\hat{\bx}_1)\mathcal{L}_{b}(\hat{\bk}\cdot\hat{\bx}_2)\mathcal{L}_{L}(\hat{\bk} \cdot \hat{\bx}_c(\bx_1,\bx_2))\,.\label{eq:generic-mapping}
\end{align}
\end{widetext}
A generic LOS, such as the mid-point LOS, can depend on both the angular and radial components of the two galaxy positions in a way that is not factorizable. As a result, the integrals in Eq.~\eqref{eq:generic-mapping} do not simplify, leading to a highly complex mapping function. Under such a generic LOS, the Yamamoto estimator receives contributions from different combinations of $\ell$ and $m$ for the two fields. In contrast, for the end-point LOS, the estimator reduces to a pseudo-$C_{\ell}$ form (\cref{eq:dSFBPS-W}) where the azimuthal mixing effects caused by window and IC are effectively ignored. 

When the LOS only depends on the angular positions of galaxies, that is $\hat{\bx}_{c}=\hat{\bx}_{c}(\hat{\bx}_1,\hat{\bx}_2)$ as in the case for the angular bisector, the angular and radial integrals of Eq.~\eqref{eq:generic-mapping} become separable
\begin{align}
    \mathcal{Q}_{n_1\ell_1m_1,n_2\ell_2m_2}^{L,ab}(k)&\sim\mathcal{V}_{n_1\ell_1}^{a}(k)\mathcal{V}_{n_2\ell_2}^{b}(k)\Omega^{L,ab}_{\ell_1m_1,\ell_2m_2}\label{eq:generic-mapping-only-angular}
\end{align}
where the radial integrals are encoded in the dSFB-to-SFB mapping function (\cref{eq:d-to-g}), and all the angular integrals produce the following coupling coefficient
\begin{align}
&\Omega^{L,ab}_{\ell_1m_1,\ell_2m_2}\equiv\int_{\hat{\bx}_1}Y_{\ell_1m_1}(\hat{\bx}_1)\int_{\hat{\bx}_2}Y^{*}_{\ell_2m_2}(\hat{\bx}_2)\nonumber\\
&\qquad\int_{\hat{\bk}}\mathcal{L}_{a}(\hat{\bk}\cdot\hat{\bx}_1)\mathcal{L}_{b}(\hat{\bk}\cdot\hat{\bx}_2)\mathcal{L}_{L}(\hat{\bk} \cdot \hat{\bx}_c(\hat{\bx}_1,\hat{\bx}_2))\,.
    \label{eq:angular-coupling}
\end{align}
We expect the explicit evaluation of this angular coupling coefficient to be computationally challenging due to the nested angular dependencies. Nonetheless, if this structure can be computed, it would enable the construction of an estimator based on the angular bisector LOS without any approximation, potentially offering a cleaner separation of the Doppler effect.

For the end-point LOS with $\hat{\bx}_{\rm c}=\hat{\bx}_1$, Eq.~\eqref{eq:angular-coupling} significantly simplifies to  
\begin{align}
\Omega^{L,ab}_{\ell_1m_1,\ell_2m_2}=\frac{(4\pi)^2}{2b+1}\begin{pmatrix}
a & L & b\\
0 & 0 & 0
\end{pmatrix}^2\delta_{\ell_1b}^{\rm K}\delta_{\ell_2b}^{\rm K}\delta_{m_1m_2}^{\rm K}\,,
\end{align}
by applying Eqs.~\eqref{eq:contraction_legendre}, \eqref{eq:LL_ortho}, and \eqref{eq:YY_ortho}. Under this simplification, the generic dSFB-to-PSM mapping in Eq.~\eqref{eq:generic-dSFB-to-PSM} reduces to the form given in Eq.~\eqref{eq:PSM-to-dSFB} with the Wigner-3$j$ symbol. Choosing the end-point LOS significantly simplifies the structure of the angular coupling coefficients in Eq.~\eqref{eq:angular-coupling}, thereby making the dSFB-to-PSM mapping more tractable and applicable in practice.

\subsection{PSM-to-SFB mapping}\label{sec:psm-to-sfb}
For completeness, we now discuss how the SFB power spectrum can be expressed as a mapping or approximation from PSM—that is, the reverse direction of the SFB-to-PSM mapping that has been the main focus of this work. This reverse mapping has been explored in Refs.~\cite{18Castorina_ZA_WA} and \cite{25ppn0} for the continuous and discrete SFB PS respectively, which we summarize here.

Assuming a uniform, full-sky survey geometry, the continuous SFB PS of Eq.~\eqref{eq:cSFBPS-def} can be related to the local PSM defined in Eq.~\eqref{eq:local-PSM} as the following \cite{18Castorina_ZA_WA}
\begin{align}
C_{l}(k_1,k_2)
&=\sum_{L,\lambda} (2\lambda+1)i^{\lambda-\ell}\tj{\ell}{L}{\lambda}{0}{0}{0}^2\nonumber\\
&\int dx\,x^2j_l(k_1x)j_{\lambda}(k_2x)P^{\rm loc}_{L}(k_2,x)\,.
\label{eq:localPSM-to-cSFB}
\end{align}
We refer readers to Section~2.4 of Ref.~\cite{18Castorina_ZA_WA} for the detailed derivation of this result. This expression demonstrates that a small number of local PS multipoles (e.g., $L=0,2,4$) are sufficient to generate all angular multipoles of the cSFB PS through some geometric coupling. We emphasize that the above expression only applies to the ideal, full-sky geometry. In practice, the exact modeling of local PSM at different redshifts is challenging, and hence Eq.~\eqref{eq:localPSM-to-cSFB} is rarely used for evaluating the SFB PS.

Applying the effective redshift approximation described in \cref{sec:ERA}, we can take the local PSM outside the radial integral by evaluating it at an effective redshift $z_{\rm eff}$ and obtain
\begin{align}
\label{eq:ClkT2}
  C_\ell(k_1,k_2) &\approx \sum_{L,\lambda} (2\lambda+1)i^{\lambda-\ell}\tj{\ell}{L}{\lambda}{0}{0}{0}^2\nonumber\\
  &\quad\mathcal{I}_{\ell\lambda}(k_1,k_2)P^{\rm loc}_L(k_2,z_{\rm eff})\,,
\end{align}
where $\mathcal{I}_{\ell\lambda}(k_1,k_2)$ is the double spherical Bessel integral defined in Eq.~\eqref{eq:double-Bessel}. This kernel is sharply peaked around $k_1=k_2$, reflecting the dominance of the diagonal components in the cSFB PS.

The discrete SFB power spectrum, which is more relevant for numerical implementation and data analysis due to its improved stability, admits an even simpler approximation. Under the plane-parallel limit and effective redshift approximation, Ref.~\cite{25ppn0} has shown that the dSFB PS can be written as
\begin{align}
     C_{\ell nn'}\approx \delta_{nn'}^{\rm K}P^{\rm loc}\left(k_{n\ell},\mu=\frac{k_{||,n\ell}}{k_{n\ell}},x_{{\rm eff},n\ell}\right)\,,
    \label{eq:pp-SFB}
\end{align}
where  $k_{||,n\ell}$ is the line-of-sight Fourier modes associated with each discrete radial basis function $g_{n\ell}(x)$, given by
\begin{align}
    k_{||,n\ell}\equiv\frac{n\pi}{x_{\rm max}-x^{\rm t}_{n\ell}}\,.
    \label{eq:k-LOS}
\end{align}
Here $x^{\rm t}_{n\ell}$ denotes the location where the radial oscillations begin, and $x_{{\rm eff},n\ell}$ is an effective radial location associated with the mode $k_{n\ell}$ for evaluating the local PS.

Equation~\eqref{eq:pp-SFB} shows that the dSFB PS can be approximated by the local power spectrum of Eq.~\eqref{eq:local-PS}. At the small angular or small radial scales, this approximation holds to percent-level accuracy under the Newtonian RSD \cite{25ppn0}. One can then expand the local PS into Legendre multipoles and obtain 
\begin{align}
     C_{\ell nn'}\approx \delta_{nn'}^{\rm K}\sum_{L}P^{\rm loc}_{L}\left(k_{n\ell},x_{{\rm eff},n\ell}\right)\mathcal{L}_{L}\left(\mu=\frac{k_{||,n\ell}}{k_{n\ell}}\right)\,,
    \
    \nonumber
\end{align}
which provides an approximation of the dSFB PS in terms of the local PSM. 

In practice, Eq.~\eqref{eq:pp-SFB} is useful for fast evaluation at the dSFB PS at small angular or small radial scales, while Eq.~\eqref{eq:localPSM-to-cSFB} provides more analytical insight into the geometric coupling between Legendre multipoles and angular multipoles.

\section{Useful Identities}

Orthogonality of spherical harmonics:
\begin{align}
    \int d^2 \hat{\bk}~ Y_{\ell m} (\hat{\bk}) Y^*_{\ell' m'}(\hat{\bk}) = \delta^{\rm K}_{\ell \ell'} \delta^{\rm K}_{m m'}\,.
    \label{eq:YY_ortho}
\end{align}

Completeness of spherical harmonics:
\begin{align}
    \sum_{\ell m}Y_{\ell m}(\hat{\bx})Y_{\ell m}^*(\hat{\bx}')=\delta^{\rm D}(\hat{\bx}-\hat{\bx}')\,.
    \label{eq:YY_complete}
\end{align}

 Orthogonality of spherical Bessel functions:
\begin{align}
\frac{2kk'}{\pi}\int_{0}^{\infty}dx\, x^2 j_{\ell}(kx)j_{\ell}(k'x)=\delta^{\rm D}(k-k')\,.
\label{eq:SB_ortho}
\end{align}
Interchanging $x$ and $k$ will give the completeness relation for spherical Bessel functions. 

Orthogonality of dSFB basis:
\begin{align}
\int d^3\bx\,e_{\mu}(\bx)e^{\nu}(\bx)=\delta^{{\rm K}}_{\mu\nu}\,.
\label{eq:dSFB-ortho}
\end{align}
Here we adopt the index notation in \cref{sec:notation-convolution}. Note that the orthogonality of the spherical harmonics in Eq.~\eqref{eq:YY_ortho} forces the discrete spherical Bessel functions to share the same $\ell$, which leads to the Kronecker delta for the radial modes.

Completeness of discrete spherical Bessel functions:
\begin{align}
\sum_{n}g_{n\ell}(x)g_{n\ell}(x')=\frac{1}{x^2}\delta^{\rm D}(x-x')\,,
\label{eq:dSB-complete}
\end{align}
for $x,x'\in[x_{\rm min},x_{\rm max}]$. The above completeness relation is guaranteed by the Sturm–Liouville theory \cite{08SL}, since the discrete spherical Bessel functions $g_{n\ell}(x)$ of a fixed $\ell$ are solutions to the radial Helmholtz equation of Eq.~\eqref{eq:helmholtz-radial} under certain boundary conditions, which forms a Sturm–Liouville problem.

Rayleigh expansion of a plane wave:
\begin{align}
    e^{i \bk \cdot \bx} &= \sum_\ell i^\ell (2\ell +1) j_\ell(kx) \mathcal{L}_\ell (\hat{\bk} \cdot \hat{\bx})\label{eq:plane_wave_exp-L}\\
    &=(4\pi)\sum_{\ell m} i^\ell j_\ell(kx) Y_{\ell m}^*(\hat{\bk}) Y_{\ell m}(\hat{\bx})\label{eq:plane_wave_exp-Y}\,,
\end{align}
where the first and the second line are related by the addition of spherical harmonics:
\begin{equation}
\label{eq:addition_spherical_harmonics}
    \mathcal{L}_\ell(\hat{\bk} \cdot \hat{\bx}) = \frac{4\pi}{2\ell +1} \sum_m Y_{\ell m} (\hat{\bx}) Y_{\ell m}^* (\hat{\bk})\,.
\end{equation}

Orthogonality of Legendre polynomials:
\begin{equation}\label{eq:LL_ortho}
    (2\ell +1) \int \frac{d^2 \hat{\bk}}{4 \pi} \mathcal{L}_\ell(\hat{\bk} \cdot \hat{\bx}) \mathcal{L}_{\ell'}(\hat{\bk} \cdot \hat{\mathbf{y}}) = \delta^{\rm K}_{\ell \ell'} \mathcal{L}_\ell(\hat{\bx} \cdot \hat{\mathbf{y}})\,,
\end{equation}
which implies that 
\begin{equation}\label{eq:integrate_L_ell}
    \int \frac{d^2 \hat{\bk}}{4\pi} e^{i \bk \cdot \bx} \mathcal{L}_\ell(\hat{\bk} \cdot \hat{\mathbf{y}}) = i^\ell j_\ell(kx) \mathcal{L}_\ell(\hat{\bx} \cdot \hat{\mathbf{y}})\,.
\end{equation}

Product of Legendre polynomials:
\begin{align}\label{eq:contraction_legendre}
    &\quad\mathcal{L}_{\ell_1} (\hat{\bk}\cdot \hat{\bx}) \mathcal{L}_{\ell_2} (\hat{\bk}\cdot \hat{\bx})\nonumber\\
    &=\sum_{\ell_3} \begin{pmatrix}
\ell_1 & \ell_2 & \ell_3\\
0 & 0 & 0
\end{pmatrix}^2 (2\ell_3+1) \mathcal{L}_{\ell_3} (\hat{\bk}\cdot \hat{\bx})\,.
\end{align}

Legendre Polynomials and Wigner-3$j$:
\begin{align}\label{eq:legendre-3j}
&\quad\int_{0}^{\pi}\mathcal{L}_{l_1}(\cos\theta)\mathcal{L}_{l_2}(\cos\theta)\mathcal{L}_{l_3}(\cos\theta)\,d(\cos\theta)\nonumber\\
&=2\begin{pmatrix}
l_1 & l_2 & l_3\\
0 & 0 & 0 
\end{pmatrix}^2\,.
\end{align}

\subsection{Higher-Order derivatives of spherical Bessel function}
We now directly prove the identity of Eq.~\eqref{eq:Legendre_SBderiv-Wigner_SB}, where the sum over spherical Bessels involving Wigner-3$j$ equals the sum over higher-order derivatives of spherical Bessels weighted by the coefficients of Legendre polynomials. Assume Legendre polynomials can be written as $\mathcal{L}_L(x)=\sum_{n=0}^{i=L}c_{Ln}x^n$ and take any constant vector $\bx$. Successively applying Eqs.~\eqref{eq:legendre-3j}, \eqref{eq:plane_wave_exp-L}, and \eqref{eq:integrate_L_ell} yields
\Needspace{12\baselineskip}
\begin{align}
&\qquad\sum_{a}i^{-a}(2a+1)\begin{pmatrix}
a & L & b\\
0 & 0 & 0
\end{pmatrix}^2 j_a(kx)\nonumber\\
&=\sum_{a}\frac{i^{-a}}{4\pi}(2a+1)j_a(kx)\int_{\hat{\bk}}  \mathcal{L}_a(\hat{\bk}\cdot\hat{\bx})\mathcal{L}_b(\hat{\bk}\cdot\hat{\bx})\mathcal{L}_L(\hat{\bk}\cdot\hat{\bx}) \nonumber\\
&=\frac{1}{4\pi}\int_{\hat{\bk}} \,e^{-i\bk\cdot\bx}\mathcal{L}_b(\hat{\bk}\cdot\hat{\bx})\mathcal{L}_L(\hat{\bk}\cdot\hat{\bx})\nonumber\\
&=\sum_{n=0}^{L}\frac{c_{Ln}}{4\pi}\int_{\hat{\bk}} \,e^{-i\bk\cdot\bx}(\hat{\bk}\cdot\hat{\bx})^n\mathcal{L}_b(\hat{\bk}\cdot\hat{\bx})\nonumber\\
&=\sum_{n=0}^{L}c_{Ln}i^{n}(-1)^b\frac{d^{n}}{d(xk)^n}\left[\frac{1}{4\pi}\int_{\hat{\bk}} d^2{\hat{\bk}}\,e^{i\bk\cdot\bx}\mathcal{L}_b(\hat{\bk}\cdot\hat{\bx})\right]\nonumber\\
&=i^{-b}\sum_{n=0}^{L}c_{Ln}i^{n}j^{(n)}_b(kx)\,.\label{eq:Flbkx}
\end{align}

\subsection{dSFB radial function under velocity boundary condition}
We aim to show that the coefficient $d_{n\ell}$ defined in Eq.~\eqref{eq:dnl} satisfies $d_{n0}\sim\delta_{n0}^{\rm K}$ under the velocity boundary condition. The dSFB radial basis functions $g_{n\ell}(x)$ satisfy the radial part of the Helmholtz equation (the Laplacian eigenequation)
\begin{align}
\frac{\dd}{\dd x}\left(x^2\,\frac{\dd g_{\ell}(kx)}{\dd x}\right)
+ \big[(kx)^2 - \ell(\ell+1)\big]\,g_{\ell}(kx)=0\,,
\label{eq:helmholtz-radial}
\end{align}

Integrating the above equation over $[x_{\rm min},x_{\rm max}]$ and considering the velocity boundary condition requires the derivatives of the basis functions to vanish at the boundary, that is $g'(kx_{\rm min})=g'(kx_{\rm max})=0$, we have
\begin{align}
    k^2 d_{n\ell}=\ell(\ell+1)\int_{x_{\rm min}}^{x_{\rm max}}dx\,g_{\ell}(kx)\,.
\end{align}
Therefore, for the SFB monopole $\ell=0$ and $n\neq 0$ such that $k_{n\ell}>0$, we have $d_{n0}=0$. 

For the DC mode with $\ell=0$ and $n=0$, we have $k_{00}=0$ under the velocity boundary condition \cite{23Gebhard_SFB_eBOSS}. One can check that Eq.~\eqref{eq:helmholtz-radial} produces a solution that is constant for $g_{00}(k_{00}x)$. The normalization condition in Eq.~\eqref{eq:gnl_orthonormality} then requires $g_{00}(k_{00}x)=(V/(4\pi))^{-1/2}$ where $V$ is volume of the spherical shell. Therefore,
\begin{align}  d_{n0}=\sqrt{\frac{V}{4\pi}}\delta_{n0}^{\rm K}\,.
    \label{eq:d0n-identity}
\end{align}

\bibliography{refs}

\end{document}